\documentclass{amsart}

\usepackage{mathtools,bm,comment,amssymb,enumitem,hyperref,cite,xcolor,array,float,subcaption,float,pgfplots}
\usepackage[capitalise,nameinlink]{cleveref}
\usepackage[english]{babel}
\usepackage[margin=1in]{geometry}
\usepackage[foot]{amsaddr}
\usepackage[linesnumbered,ruled]{algorithm2e}
\usepackage[noend]{algpseudocode}
\usepackage{textcomp}
\usepackage{placeins}
\usetikzlibrary{shapes,arrows}
\usepackage[textsize=tiny]{todonotes}
\usepackage{units}
\usepackage{amsmath}
\usepackage{xurl}
\usepackage{multirow}
\usepackage{graphicx}
\graphicspath{{Images/}{../}}

\numberwithin{equation}{section}
\crefname{subsection}{Subsection}{Subsections}
\crefformat{equation}{(#2#1#3)}
\crefformat{enumi}{(#2#1#3)}
\crefname{figure}{Figure}{Figures}

\renewcommand{\t}{\tilde}

\newcommand{\R}{\mathbb{R}}

\newcommand{\vtheta}{\boldsymbol{\theta}}

\DeclareMathOperator{\sign}{sign}

\newcommand{\argmin}{\operatornamewithlimits{argmin}}

\makeatletter
\@namedef{subjclassname@2020}{\textup{2020} Mathematics Subject Classification}
\makeatother

\makeatletter
\renewcommand{\email}[2][]{%
	\ifx\emails\@empty\relax\else{\g@addto@macro\emails{,\space}}\fi%
	\@ifnotempty{#1}{\g@addto@macro\emails{\textrm{(#1)}\space}}%
	\g@addto@macro\emails{#2}%
}
\makeatother

\makeatletter
\pgfmathdeclarefunction{erf}{1}{%
	\begingroup
	\pgfmathparse{#1 > 0 ? 1 : -1}%
	\edef\sign{\pgfmathresult}%
	\pgfmathparse{abs(#1)}%
	\edef\x{\pgfmathresult}%
	\pgfmathparse{1/(1+0.3275911*\x)}%
	\edef\t{\pgfmathresult}%
	\pgfmathparse{%
		1 - (((((1.061405429*\t -1.453152027)*\t) + 1.421413741)*\t 
		-0.284496736)*\t + 0.254829592)*\t*exp(-(\x*\x))}%
	\edef\y{\pgfmathresult}%
	\pgfmathparse{(\sign)*\y}%
	\pgfmath@smuggleone\pgfmathresult%
	\endgroup
}
\makeatother

\let\originalleft\left
\let\originalright\right
\renewcommand{\left}{\mathopen{}\mathclose\bgroup\originalleft}
\renewcommand{\right}{\aftergroup\egroup\originalright}

\newcommand{\uR}{10^8\;\unit{Pa}\;\unit{s}\;\unit{m}^{-3}}
\newcommand{\uC}{10^{-8}\;\unit{m}^{3}\;\unit{Pa}^{-1}}

\title[Calibration of Windkessel parameters]{Data-Driven Neural Networks for Windkessel Parameter Calibration}

\author{Benedikt Hoock$^{1,*}$, Tobias K\"oppl$^{2,3}$}
\thanks{${}^*$Corresponding author}
\email{benedikt.hoock@tum.de}
\email{tobias.koeppl@fokus.fraunhofer.de}

\begin{document}
\maketitle
\vspace*{-2mm}
\begin{center} \footnotesize
	$^1$School of Computation, Information and Technology, Technical University of Munich, Germany \\
	$^2$Faunhofer FOKUS Institute, System Quality Center (SQC), Berlin, Germany 
    \\
    $^3$Faculty of Science, Hasselt University, Belgium
\end{center}

\begin{abstract}
In this work, we propose a novel method for calibrating Windkessel (WK) parameters in a dimensionally reduced 1D-0D coupled blood flow model. To this end, we design a data-driven neural network (NN)trained on simulated blood pressures in the left brachial artery. Once trained, the NN emulates the pressure pulse waves across the entire simulated domain, i.e., over time, space, and varying WK parameters, with negligible error and computational effort. To calibrate the WK parameters on a measured pulse wave, the NN is extended by dummy neurons and retrained only on these. The main objective of this work is to assess the effectiveness of the method in various scenarios -- particularly, when the exact measurement location is unknown or the data are affected by noise.
\end{abstract}

\vspace{0.2cm}
\vspace{0.1cm}

\section{Introduction}
Numerical simulation of blood flow has steadily gained interest in recent decades, driven by advances in computational power, efficient numerical algorithms, and imaging and reconstruction techniques. These developments are motivated by the fact that mathematical models enable clinical doctors and physiologists to study cardiovascular diseases non-invasively \cite{schwarz2023beyond,ambrosi2012modeling,Quarteroni,koeppl2018numerical,koppl2023dimension,quarteroni2000computational}.

An important class of mathematical models commonly applied in this context are 1D-0D coupled blood flow models \cite{alastruey2007modelling,alastruey2012physical,chen2013simulation,mynard20081d,mynard2015one,gyurki2024central}, combining one-dimensional (1D) Navier-Stokes equations for larger vessels with lumped parameter 0D models \cite{alastruey2008lumped,ghitti2022nonlinear} for the smaller vessels. The 0D models, formulated as ordinary differential equations (ODEs) without spatial variable, express the resistive and compliant effects -- collectively known as the Windkessel (WK) effect -- and are thus referred to as Windkessel models.

In this work, we consider the 3-element WK model \cite{alastruey2008lumped}, which comprises two resistance parameters, representing the terminal vessel and the connected smaller vessels downstream, and one compliance parameter, reflecting the blood storage capability of the omitted vessels. Accurate simulation depends on a careful calibration of the WK parameters for each outlet. A standard approach sets one resistance parameter to the hydraulic resistance of the terminal vessel \cite{formaggia2006numerical}, while the remaining parameters are determined by distributing the total resistance and compliance of the arterial system among the different WK models at the outlets. Thereby, the compliance parameter and the sum of the resistance parameters, known as peripheral resistance, are determined. From the peripheral resistance and the precomputed first resistance, the second resistance parameter can be obtained. The distribution of the total resistance and compliance to the individual parts is based on principles from electrical science \cite[Section 2.2]{alastruey2007modelling}\cite{alastruey2008lumped,renner2024accelerated}. However, this method lacks the involvement of patient-specific measurements, which is essential for personalized modeling. An alternative is to formulate an optimization problem that minimizes discrepancies between measured and simulated data to yield the calibrated parameters. To find these, both classical optimization methods \cite{du2023personalized,fevola2022vector,ismail2013adjoint} and statistical methods, such as the Bayesian calibration method \cite{spitieris2022bayesian,olsen2022bayesian,salvador2023fast,richter2025bayesian}, have been studied.

In this work, we propose a calibration method that combines the standard parameter-distributing approach with a classical optimization procedure, relying on blood pressure measurements, specifically in the left brachial artery -- a common site to monitor the arterial blood pressure in a non-invasive way \cite{jamison2022brachial,samartkit2024non,zhou2024clinical}. For generality, we use synthetic data simulated by the 1D-0D coupled blood flow model as proxies for the required measurements, optionally disturbed by random noise. Our design application, however, is real-time calibration using continuous, ideally non-invasive, arterial blood pressure data -- the gold standard of clinical blood pressure measurement -- such as those obtained from wearable ultrasound sensors developed very recently\cite{zhou2024clinical,min2025wearable}. Our primary goal is to calibrate the total resistance and compliance of the 1D-0D model, based on mean-square error between measured and simulated pressures. These global WK parameters are then distributed to the site-specific WK parameters of the different outlets by the standard method. The optimization also infers the location of the measurement device along the brachial artery -- to account for uncertainty in sensor placement -- and includes a phase shift parameter to synchronize the measured and simulated data. 

A similar approach was explored in \cite{ventre2021parameter}\cite[Section 8.4]{ventre2020reduced}, where the WK parameters of a purely 0D model were calibrated by minimizing a quadratic objective function with respect to invasively measured blood pressures in a brachial artery. The resulting WK parameters were then used to compute the global resistance of the 0D model. Finally, the global resistance of a 1D-0D model is calibrated by minimizing its quadratic difference to the found global resistance of the 0D model. In \cite{richter2025bayesian}, a similar strategy is used to determine the WK parameters of a 3D-0D coupled blood flow model. Thereby, a purely 0D model with already calibrated parameters is used to compute a posterior distribution of the WK parameters for a 3D-0D blood flow model. Both outlined approaches rely on computationally inexpensive 0D models to estimate the parameters of a higher-dimensional model.

To enable a fast and accurate calibration in our setting, we introduce a neural network that acts as a surrogate model for the computationally expensive 1D-0D simulations at the brachial artery. This surrogate is a fully connected feedforward neural network (FCFNN) trained to replicate the map from time, position, and global WK parameters to the simulated blood pressure. The training data are generated by the coupled 1D-0D model, which can produce a fine-resolved and representative reference set thanks to efficient implementation and dimension reduction. Unlike physics-informed neural networks (PINNs), which integrate residuals of the governing equations into training \cite{cai2021physics,raissi2019physics,kissas2020machine}, our method works with a fully data-driven and parametrically sparse neural network. This is viable, since our reference data provides sufficient information on the system, and avoids the complexity in the neural network inherent many PINN applications. After training, we extend the architecture of our FCFNN by dummy neurons for the calibration variables -- the WK parameters, the position, and the phase shift. Only these are optimized during retraining to match a given measured pulse wave, yielding the calibrated parameters as well as the optimal position and phase.

The remainder of this paper is structured as follows: Section \ref{sec:bloodflow} presents the 1D-0D coupled blood flow model and the numerical solution techniques. Section \ref{sec:calibration} describes the calibration process for the WK parameters. Next, Section \ref{sec:the_NN} details the training of the NN surrogate and its modification for the actual calibration. Section \ref{sec:tests} demonstrates the performance of our approach in a series of numerical tests. Finally, Section \ref{sec:conclusions} concludes the paper and outlines directions for future work.
\section{Dimensional reduced modeling of blood flow}
\label{sec:bloodflow}
\begin{figure}[h!]
	\centering
	\includegraphics[width=0.99\textwidth]{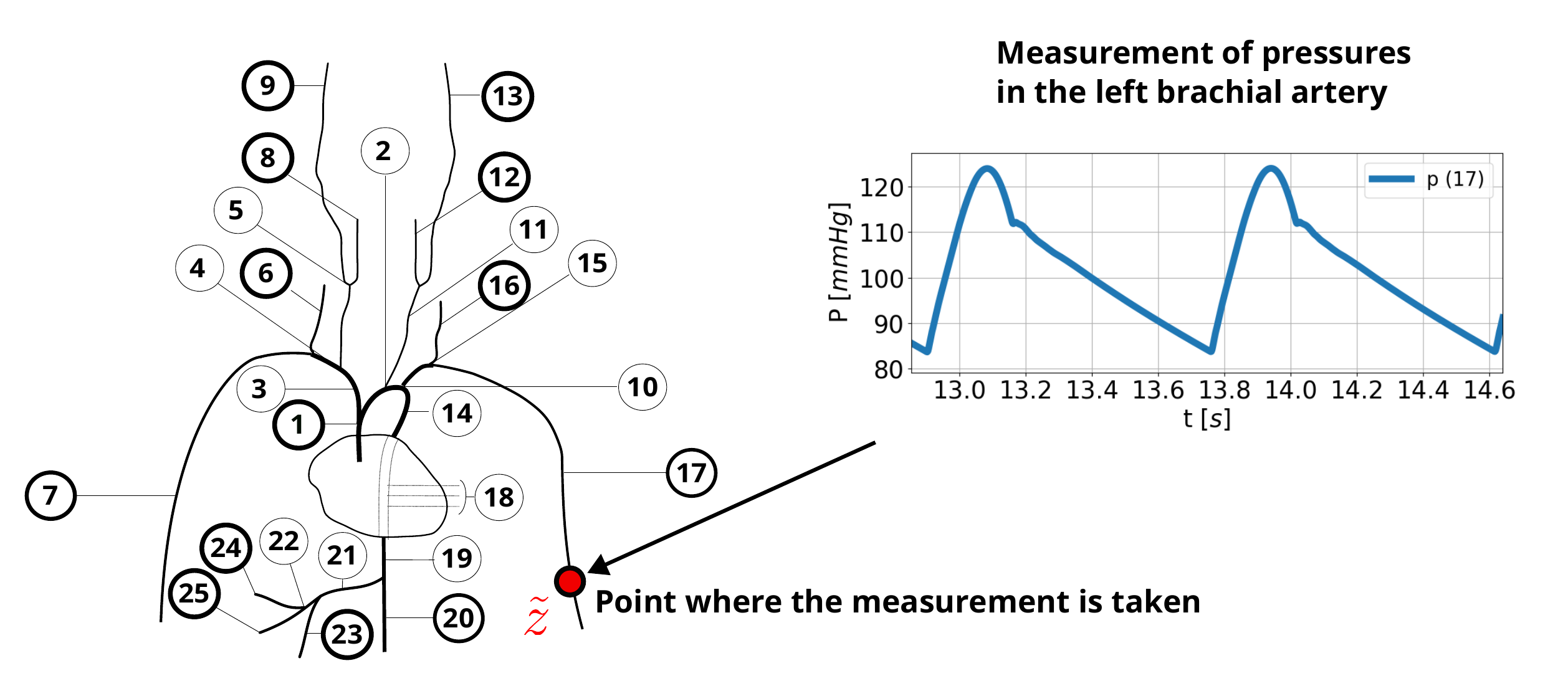}
	\caption{\label{fig:Network} Schematic representation of the studied network, composed of the large arteries of systemic circulation. The 25 considered vessels are numbered as indicated, which is also used in Tab. \ref{tab:DataNetwork}. The network contains the aorta, the carotid arteries, the hepatic arteries, and the left and right brachial arteries. The calibration procedure assumes that blood pressures are measured at some point $\tilde{z}$ within the left brachial artery (Vessel $17$).}
\end{figure}

In this work, we consider a network of larger systemic arteries, including the aorta, its branches connecting the liver, both carotid arteries, and both brachial arteries (see Figure \ref{fig:Network}). Networks of this type are useful for estimating the blood volume distribution across different body regions. The studied network can estimate the blood supply of the liver -- if calibrated appropriately, also on a patient-specific level. This can improve predicting the blood perfusion of the liver and thus help in planning heat therapy for the treatment of liver cancer \cite{sharma2022current}.

The physiological properties of the vessels are important model parameters of our network. We used the average lengths, radii, wall thicknesses, and elasticity parameters as reported in Table \ref{tab:DataNetwork}, which are adopted from \cite{alastruey2007modelling,stergiopulos1992computer}. A 1D-0D coupled model \cite{peiro12004numerical} is employed to simulate the blood flow in this network. This model is based on a domain decomposition approach: first, the network is split into individual vessels; then, a simplifying 1D flow model is assigned to each vessel; finally, the different flow models are coupled at their interfaces to simulate the blood flow in the entire network.
In the remainder of this section, we present the core principles of the 1D model and the coupling conditions at bifurcations. Moreover, we discuss the zero-dimensional 3-element Windkessel (WK) model, which represents the omitted parts of the vascular tree in our model. Lastly, we describe the numerical solution methods -- these are applied to generate the data for the neural network. All model parameters, except for the WK parameters, are summarized in the appendix. 

\subsection{Modeling of blood flow through a single vessel}
\label{subsec:1D}
All of the 25 larger arteries of the cardiovascular network are described one-dimensionally, i.e., as cylindrical tubes of a certain length $l$ (the vessel index $i$ is omitted for brevity), with a deformable wall and aligned along the $z$-axis. Blood is treated as an incompressible and Newtonian fluid, which is reasonable for large to medium-sized arteries. Arterial wall deformation is assumed to occur only in the radial direction, blood flow is considered as radially symmetric, where the $z$-component of the velocity field is dominant. Under these assumptions, integrating the Navier-Stokes equations yields the following system of partial differential equations (PDEs), essentially describing conservation of mass and momentum in 1D \cite{vcanic2003mathematical,hughes1973one,koppl2023dimension}:
\begin{align}
\label{eq:1D_PDEs} 
	\frac{\partial A}{\partial t} + \frac{\partial Q}{\partial z} = 0, \quad\quad \frac{\partial Q}{\partial t} + \frac{\partial}{\partial z}\left( \frac{Q^2}{A} \right) + \frac{A}{\rho}\frac{\partial p}{\partial z} = -K_{r} \frac{Q}{A}.
\end{align}
Here, $A=A(z,t)$, $Q=Q(z,t)$ and $p(z,t)$ represent the cross-sectional area, volume flow and pressure at position $z$, time $t>0$ and within a certain vessel. $\rho$ is the density of blood and $K_r = 22 \pi \mu$ is a resistance parameter depending on the kinematic viscosity of blood $\mu\;\left[\unitfrac{cm^2}{s} \right]$ \cite{ventre2020reduced}[Chapter 8.4.2].

The PDE system \ref{eq:1D_PDEs} is closed by an algebraic equation derived from the Young-Laplace equation \cite{olufsen1999structured}\cite{koppl2023dimension}[Chapter 3.2](FSI 4): 
\begin{equation}
	\label{eq:PressureAreaRelation}
	p(z,t) = G_{0} \left( \sqrt{\frac{A}{A_{0}}} - 1 \right),\quad G_{0} = \frac{\sqrt{\pi} \cdot h_{0} \cdot
		E}{\left(1-\nu^2\right) \cdot \sqrt{A_{0}}}
\end{equation}
where $E$, $A_{0}$, and $h_{0}$ denote the Young modulus, cross-sectional area at rest, and wall thickness at rest, which are three vessel-specific constants. $A_{0}$ is obtained from the vessel's equilibrium radius via $A_{0} = R_{0}^2 \cdot \pi$. $\nu=0.5$ is the Poisson ration, a valid choice for incompressible biological tissue. $G_0$ connects the vessel-specific constants and can be interpreted as a characteristic pressure. Note that the pressure-area relation of Equation \eqref{eq:PressureAreaRelation} assumes instantaneous equilibrium of the vessel wall and the forces acting on it. Omitted effects like viscoelasticity can, e.g., be accounted for by a differential pressure law \cite{devault2008blood,mynard2015one,valdez2009analysis}. 

\subsection{Modeling of bifurcations and heartbeats}
The arterial system exhibits several levels of branching when moving downstream through the body. This corresponds to bifurcations between adjacent tubular 1D parts modeled by PDEs of the type of \ref{eq:1D_PDEs}. Bifurcations must be described accurately to produce reliable simulations, which is why their mathematical formulation has been studied extensively \cite{carson2017implicit,formaggia2003one,holden1999riemann,koppl2013reduced}. In our system, the adjacent 1D parts are coupled by enforcing mass conservation and continuity of the total pressure at the points of the bifurcation.

The driving term of the blood dynamics is given by the pulsation of the heart, which mathematically imposes a boundary condition on the inlet of Vessel $1$ at its connection to the heart, i.e. at position $z_1=0$. We model this by the flow rate profile
\begin{equation}
	\label{eq:Qheart}
	Q_1\left(0,t\right) = Q_{\max} \cdot
	\begin{cases}
		\sin\left( \frac{3 \cdot \pi \cdot t}{T} \right),\; 0 \leq t \leq \frac{T}{3}, \\
		\\
		0.0, \; \frac{T}{3} < t \leq T.
	\end{cases}
\end{equation}
Here, $Q_{\max}\;\left[\unitfrac{cm^3}{s}\right]$ is the maximal flow rate and $T\;\left[\unit{s}\right]$ is the duration of a single heartbeat, and the profile is extended periodically.

\subsection{Modeling of the omitted vessels}
\label{sec:omitted}
To account for the effect of the vessels not explicitly modeled in the network of Figure \ref{fig:Network}, we use a lumped parameter model to approximate the behaviour in the omitted parts. This formulates an ordinary differential equation (ODE) for each outlet to the omitted part which is coupled to the respective 1D part. Since the ODEs do not depend on the spatial dimension, they are referred to as zero-dimensional (0D) models. A common lumped-parameter model is the Windkessel model\cite{Quarteroni}, which draws an analogy to an electrical circuit with resistive and capacitive parts. The former accounts for the counter-pressure of the omitted vessels, the latter for the vascular storage capability.

In this scope, the term WK effect describes the maintenance of a continuous blood supply of organs and tissue resulting from the storage of a certain amount of blood volume in the 'reservoir' of the large deformable vessels \cite[Section 1.1.2]{koppl2023dimension}. The model is idealizing since the compliant and resistant effects do not separate very strictly into the large and the omitted vessels, since many of the omitted vessels exhibit some storage capacity, and the larger vessels may impose some resistance. Nevertheless, the WK model is able to produce realistic blood pressure curves, in particular, having the potential to capture the typical amplitudes and the characteristic slow decay during the diastole of the blood pressure \cite{alastruey2008lumped}.

\begin{figure}[h!]
	\centering
	\includegraphics[width=0.50\textwidth]{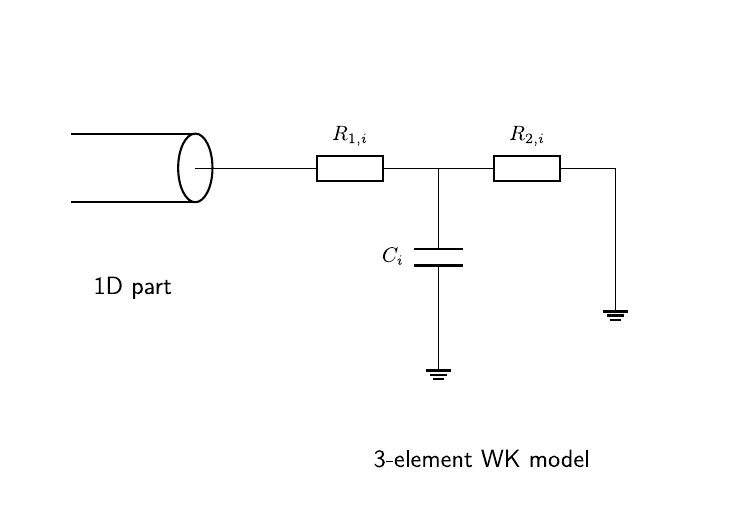}
	\caption{\label{fig:coupling_1D0D} The coupling of the 1D part to the three-element Windkessel model, which is analogous to an electrical circuit of two resistors and a capacitor.}
\end{figure}

In this work, we apply a WK model variant with three parameters -- the \emph{three-element WK model} \cite{alastruey2008lumped,stergiopulos1992computer} -- which correspond to a capacitor and two resistors (see Figure \ref{fig:coupling_1D0D}). For each terminal vessel $i$, the model assigns a peripheral resistance that splits into two parts:
$$
R_{p,i} = R_{1,i} + R_{2,i}.
$$
Here, $R_{1,i}$ represents the \emph{arterial} resistance adjacent upstream to the outlet of $i$, and $R_{2,i}$ the \emph{peripheral} resistance for the arterioles and capillaries downstream connected to $i$ \cite{alastruey2008lumped}. These two parameters incorporate the reflections of the pulse waves at the subsequent vessels to the model. The third parameter is the compliance $C_i$, quantifying the blood storage capacity of the omitted vessels. The triplet $\left(R_{1,i},C_i,R_{2,i} \right)$ is usually simply referred to as WK parameters \cite[Chap. 10]{Quarteroni}.

The governing ODE of the three-element WK model is
$$
p_{i,T} + R_{2,i} C_i  \frac{d p_{i,T}}{dt} = p_v + \left( R_{1,i} + R_{2,i} \right)  Q_{i,T} + R_{1,i} R_{2,i} C_i \frac{d Q_{i,T}}{dt},
$$
obtained by averaging techniques and an analogy from electrical science \cite{alastruey2007modelling,marchandise2009numerical,fevola2022vector}. It relates the pressure $p_{i,T} = p\left(A_i\left(l_i,t\right)\right)$ to the flow rate $Q_{i,T} = Q_i\left(l_i,t\right)$ at the outlet $z=l_i$ of a terminal vessel $i$ where $p_v$ denotes the average pressure in the venous system.

One observes that, after rearranging the equation, $dp_{i,T}/dt$ depends, among other variables and parameters, inversely on $C_i$, such that large values of the compliance tend to "damp" the dynamics in the pressure, and vice versa, smaller ones are accompanied by pressure curves with steeper slopes. In a medical context, high compliance is linked with vessels whose vessel walls exhibit a low stiffness. Such vessels can potentially store a higher amount of blood, while pathologically stiffer ones tend to are filled up earlier. This results in larger pressure amplitudes, which can cause vascular diseases. The two resistance parameters both have an almost linear effect on the blood pressure, but affect different phases in the cardiac cycle: The arterial resistance $R_{1,i}$ mainly scales the systolic peak, i.e., the immediate reaction to inflow, while the peripheral resistance $R_{2,i}$ acts as a global shift, lifting or dropping the pressure curve.

Now, the question of how to determine the parameters $R_{1,i}$, $R_{2,i}$ and $C_i$ arises. Possible choices for the arterial resistances $R_{1,i}$ can be found in literature \cite{alastruey2007modelling,alastruey2008lumped,chen2013simulation}. In this work, we follow \cite{alastruey2007modelling} and identify the $R_{1,i}$ with the characteristic impedance of vessel $i$, given by
$$
R_{1,i} = \frac{\rho \sqrt{\frac{G_{0,i}}{2 \rho}}}{A_{0,i}}.
$$
This choice minimizes the reflections of pressure and velocity waves that cross the outlet of $i$.

The remaining parameters $R_{2,i}$ and $C_{i}$ have to be found for all terminal vessels in the network (Figure \ref{fig:Network}), indexed by the set
$$
I_{out} = \left\{ 6,7,8,9,12,13,16,17,18,20,23,24,25\right\}.
$$
This comprises 13 terminal vessels, resulting in 26 WK parameters to be determined.

To this extent, we adapt a calibration method described in \cite{alastruey2007modelling}[Section 2.2]. It assumes that the entire systemic arterial network can be characterized by a total resistance $R_{tot}$ and a total compliance $C_{tot}$ which satisfy the relations
\begin{equation}
   \label{eq:def_Rtot_Ctot}
   \frac{1}{R_{tot}} \approx \sum_{i \in I_{out}} \frac{1}{R_{p,i}} \quad \text{ and } \quad
C_{tot} = \sum_{i=1}^{13} C_{net,i} + \sum_{i \in I_{out}} C_i.
\end{equation}
This follows from Kirchhoff's laws for parallel resistors and capacitors in an electrical circuit. The approximation in the first relation stems from neglecting the resistance of the arterial network, i.e., the inner part of Fig. \ref{fig:Network}, which is permissible, since the flow resistance in larger arteries is relatively low.

As already pointed out, the three-element WK does not clearly separate inner and peripheral contributions. In particular, the compliances $C_{net,i}$ of the arteries modeled in the 1D parts (those shown in Figure \ref{fig:Network}) must be accounted for, as their large diameters and high deformability cause significant blood storage. Following \cite{alastruey2007modelling}[Section 2.2]\cite[Section 3.6.2]{koppl2023dimension}, these can be determined by:
$$
C_{net,i} = \frac{2 \cdot l_i \cdot A_{0,i}}{G_{0,i}}.
$$
Experimental determination of these parameters would require measuring the geometry and pulse wave speed \cite{alastruey2008lumped}. Here, we instead use a simplified assumption about the distribution of the relative geometries, which allows us to assign proportions of the total resistance $R_{tot}$ and total compliance $C_{tot}$ to the individual peripheral resistances and compliances. Specifically, we allocate $20\%$ of the inverse peripheral resistances and total compliance to outlets supplying the head, indexed by
$$
I_{head} = \left\{6,8,9,12,13,16\right\} \subset I_{out}.
$$
This proportion reflects the distribution of the cardiac output within the systemic arterial system. Another $5\%$ is assigned to the left and right arms,
$$
I_{arm,r} = \left\{7\right\} \subset I_{out} \text{ and } I_{arm,l} = \left\{17\right\} \subset I_{out}.
$$
The remaining $70\%$ are assigned to outlets perfusing the rest of the body \cite[Section 2.2]{alastruey2007modelling}:
$$
I_{body} = \left\{ 18,20,23,24,25 \right\}.
$$
For subsystem $k \in \left\{head;\;arm,r;\;arm,l;\;body\right\}$, the corresponding peripheral resistance and compliance hence are given by
$$
\frac{1}{R_{p,k}} = r_k \cdot \frac{1}{R_{tot}},\;C_{p,k} = r_k \cdot C_{p,tot},
$$
i.e. as proportions of $\frac{1}{R_{tot}}$ and $C_{p,tot} = C_{tot}-\sum_{i=1}^{13} C_{net,i}$, where the allocation factors $r_k$ are given by the perfusion rates $\left\{ 20\%, 5\%, 5\%, 70\% \right\}$ derived above.

Next, these resistances and compliances of the subsystems are distributed to the respective outlets connected to those parts. For the two arms, this is trivial since they possess only a single outlet. For $body$ and $head$, the distribution to the outlets is done proportionally to the relative sizes of the cross-sectional areas:
$$
\frac{1}{R_{p,i}} = \frac{A_{0,i}}{\displaystyle{\sum_{j \in I_k} A_{0,j}}} \cdot \frac{1}{R_{p,k}} \text{ and }
C_{p,i} =  \frac{A_{0,i}}{\displaystyle{\sum_{j \in I_k} A_{0,j}}} \cdot C_{p,k},\;
i \in I_k,\;k \in \left\{head,body\right\}.
$$
This finally completes the rules for obtaining all individual Windkessel parameters from the total resistance and compliance. Thus, once these total parameters have been found by a calibration procedure, the entire 1D-0D cardiovascular model is parametrized with reasonable values.

\subsection{Numerical solution techniques}
\label{subsec:solver}
While we will later develop a surrogate model to replace the 1D-0D model, this surrogate will rely on high-quality data from explicit simulations using the 1D-0D model itself. In these computations, the PDE-system \eqref{eq:1D_PDEs} is solved by the numerical method of characteristics (NMC), a quasi-explicit and low-order method. Its first-order convergence in space and time has been shown in \cite[Theorem 1]{Riviere}. To mitigate large dissipation and dispersion errors \cite{quarteroni2006numerical}, we sample the space and time on a sufficiently fine grid. Since our models are one-dimensional, sampling on a fine spatial grid is computationally feasible. Explicit methods require also small time-step cases in that case, to provide numerically stable results. Here, the NMC has the advantage that its time step sizes are not bounded by a CFL condition \cite[Proposition 2]{Riviere}. Thus, the NMC can be run efficiently on a fine spatial grid if the time-step size is only small enough to resolve the convection-dominated blood flow, but not excessively small. In Section \ref{subsec:data}, we describe more computational details of the simulated reference data.

\section{Calibration of the Windkessel parameters}
\label{sec:calibration}
It has been pointed out that the 1D-0D method requires an accurate parametrization of its Windkessel (WK) parameters to provide reliable blood flow simulations, and that the WK parameters of its individual outlets can be obtained via the total peripheric resistance $R_{tot}$ and the total compliance $C_{tot}$. In the following, we describe a general strategy to calibrate these two global WK parameters based on measured blood pressures. For brevity, we will use the notation $R$ and $C$ to refer to the two calibration variables $R_{tot}$ and $C_{tot}$ as defined by \eqref{eq:def_Rtot_Ctot}.
The measurement, which is the reference for the calibration, is given by a time series of blood pressures, taken at a fixed position along the left brachial artery (Vessel $17$ in Fig. \ref{fig:Network}). We assume it covers exactly one cardiac cycle $I_{heart}$:
$$
\tilde{p}_{17}\left( t_{m,j}\right),\;t_{m,j} \in I_{heart}, \; j \in \left\{1,\ldots,M\right\},
$$
recorded at $M$ equidistant times $t_{m,j} \in I_{heart}$, and that its exact position $\tilde{z}$ within Vessel $17$ is recorded (we mark measurements by a tilde).

The goal is to find the optimal model parameters $R_{opt}$ and $C_{opt}$ such that the numerical prediction at $\tilde{z}$ matches the measured series $\tilde{p}_{17}\left(t_{m,j}\right)$ as close as possible. This is formulated as the minimization problem
$$
\left( R_{opt},C_{opt} \right) = \argmin_{\left(R,C\right) \in Q_2} J\left(R,C\right),
$$
where $J\left(R,C\right)$ is a cost functional and $Q_2 = \left[R_{\min},R_{\max}\right] \times \left[C_{\min},C_{\max}\right]$ the domain. In this work, we define the cost functional as the mean squared error (MSE):
\begin{equation}
	\label{eq:costfun}
	J\left(R,C\right) = \frac{1}{M} \sum_{j=1}^M \Big( \tilde{p}_{17}\left(t_{m,j}\right)- p_{17}\left(\tilde{z},t_{m,j},R,C\right) \Big)^2,
\end{equation}
between the measurement and the simulation. This idea of a least-squares optimization is straightforward and has been applied for calibration of WK parameters, e.g., in \cite{romarowski2018patient,ventre2020reduced}.

To account for the modalities of realistic measurements, the minimization problem must be confined to two aspects: (i) spatial uncertainty and (ii) temporal misalignment. Regarding the first, we notice that calculating $J$ requires evaluating the model at the correct position $\tilde{z}$ since it captures the spatial dependency of the blood pressure in Vessel $17$. A change of $z$ mainly affects the amplitude of the pulse wave, albeit the effect along the artery is small. But if the model assumes a wrong position for the measurement, in the absence of a precise position measurement, the calibration accuracy is reduced.

\begin{figure}[h!]
	\centering
	\includegraphics[width=0.85\textwidth]{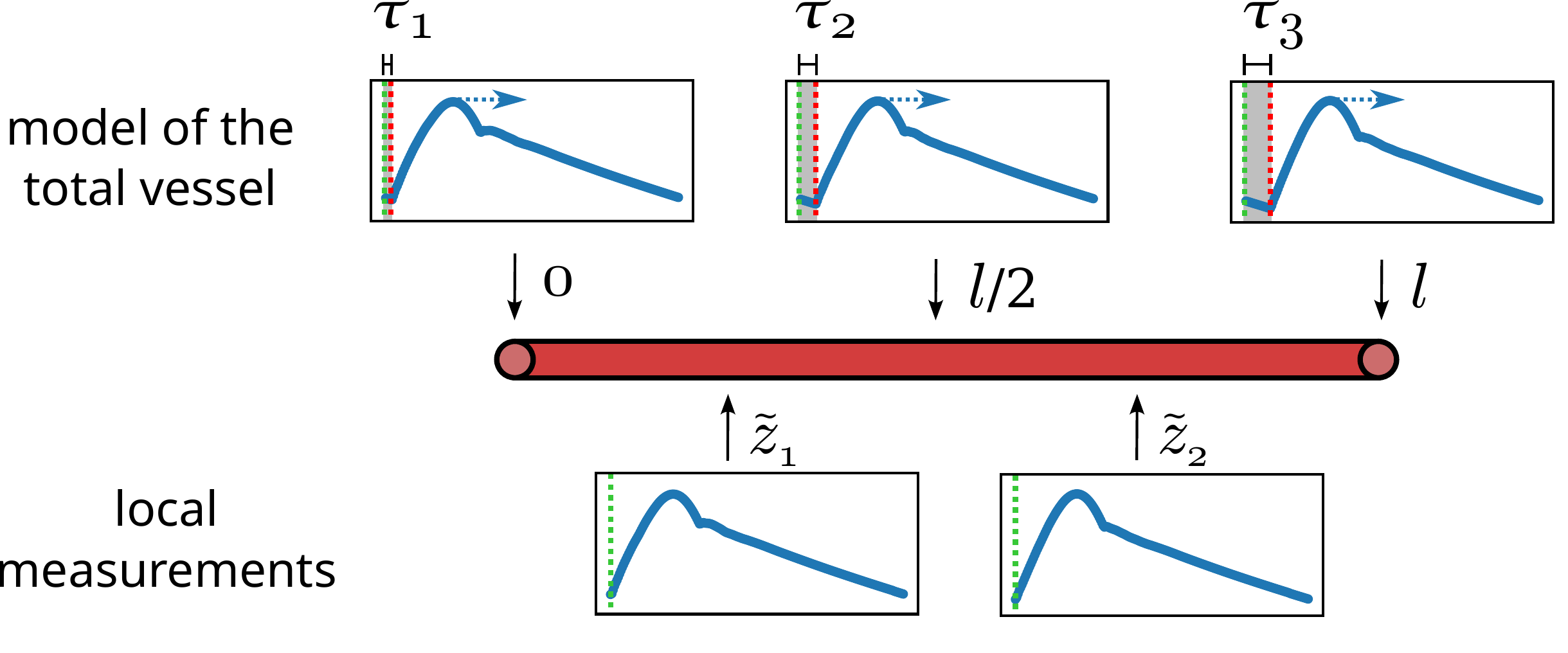}
    	\caption{\label{fig:phase_shift} Illustration of the phase shift between measurements and the model. Top: pressure profiles over one period at different positions of the vessel, as calculated by the 1D-0D model. This considers the global dynamics along the whole artery in a correlated way, including the propagation of the pulse wave through the vessel with the phase velocity (blue dotted arrows at the maximum). This leads to a phase shift of the waves along the vessel, represented by a phase-lag $\tau$ -- marked by the grey areas between $t=0$ (green lines) and the time of the minimum pressure (red lines). The phase-lag increases when moving from the inlet to the outlet. Bottom: pressure profile if measured by a local apparatus. This will put the phases to the same time points, independent of the position. By convention, the measurements could start, e.g., at the minimum, i.e., the diastolic pressure (green lines).}
\end{figure}

Regarding the second, one has to ensure that the measured and simulated pressures in \ref{eq:costfun} are synchronous in time. This could be violated by a different convention of the time zero in measurement and simulation. Moreover, the model captures the pulse propagation through the vessel, as illustrated in Fig. \ref{fig:phase_shift}, resulting in a phase lag $\tau$ that increases with the distance from the inlet. This effect is resolved since the PDEs of \ref{eq:1D_PDEs} consider the spatio-temporal behavior in the vessel at once. In contrast, a local measurement device is blind to this effect and typically will put the signal start to the same feature of the pulse wave, e.g., the diastolic pressure, regardless of the position. In total, this may result in a temporal mismatch between the measurement and the simulation.

To correct these two shortcomings, we extend the cost functional by the position $z$ and a phase-shift parameter $\tau$:
\begin{equation}
\label{eq:costfun4p}
J\left(R,C,z,\tau\right) = \frac{1}{M} \sum_{j=1}^M \Big( \tilde{p}_{17}\left(t_{m,j}\right)- p_{17}\left(z,t_{m,j}+\tau,R,C\right) \Big)^2.
\end{equation}
The minimization problem then becomes
\begin{equation}
	\label{eq:minprobtau}
	\left( R_{opt},C_{opt}, z_{opt}, \tau_{opt} \right) = \argmin_{\left(R,C,z,\tau\right) \in Q_4} J\left(R,C,z,\tau\right),
\end{equation}
where the variables are constraint to $Q_4 = Q_2 \times [0,L] \times [0,T]$. This joint optimization infers the most plausible WK parameters, with inferring the best spatio-temporal match of the simulation with the measurement.

Technically, the minimizer of $J$ can be found by means of an iterative optimization method, requiring simulating $p_{17}$ with the 1D-0D coupled blood flow model of Section \ref{sec:bloodflow} at every update step -- despite efficient numerical methods and dimensionality-reduction, this is computationally impractical for "on-the-fly" optimization. To overcome this issue, we replace the explicit simulation by a less expensive surrogate model $\hat{p}_{17}$. This surrogate $\hat{p}_{17}$ must accurately reproduce the behavior of the 1D-0D blood flow model $p_{17}$ on the entire domain of the minimization, i.e. it has to be ensured that
$$
\hat{p}_{17}(z,t,R,C) \approx p_{17}(z,t,R,C) \quad \forall (z,t,R,C) \in Q_4.
$$
The following section will detail the construction and parametrization of this surrogate model using a simple and small-sized neural network trained on a reference data set.
\section{Constructing the Surrogate as a Data-Driven Neural Network}
\label{sec:the_NN}
The surrogate model, required for an efficient calibration of the WK parameters, is constructed as a fully connected feedforward neural network (FCFNN) and trained in a purely data-driven way on blood pressure data at Vessel 17, computed explicitly by the 1D-OD beforehand. In this section, we first justify the choice of an FCFNN for the modeling of pulse waves from a broader theoretical perspective, and then describe the basic structure of the proposed NN, its training procedure, and the reference data set. Next, a series of numerical studies is presented that concerns the hyperparameters of the model, i.e., its topology in terms of the network depth and width, and the type of activation functions. This leads to an optimal version of the NN, which is assessed in its predictive qualities. In addition, we discuss the sensitivity of the surrogate on the resolution in the reference data. Lastly, we present a method for how the NN can be enhanced to solve the minimization problem directly by retraining it on a specific measurement.

\subsection{Justification of the approach}
\label{sec:justification}
Neural networks (NNs) have spread across scientific disciplines as a powerful and flexible machine learning technique that can be trained and evaluated efficiently, thanks to many high-level implementation frameworks. Currently, NNs are the standard tool to build a mathematical model for a complex relation between multiple independent and dependent variables from reference data, if the focus is more on numerical accuracy than on interpretability. Specifically in the field of cardiovascular flow simulations, there have been several successful use cases for NNs \cite{nolte2022inverse}\cite[Section 3.2]{pfaller2024ROM}. While many of the recent approaches to some extent tailor the NNs to the problem, our work is based on a standard feed-forward neural network, trained in a purely data-driven way on reference data. The theoretical foundation for this approach is given by the universal approximation properties of NNs.

Recall that a multilayer fully connected feedforward neural network (FCFNN) represents a function from $\R^n$ to $\R$ by multiple connected stacked dense layers of a certain number of artificial neurons, as shown in Figure \ref{fig:NN_enhanced} on the top. Mathematically, the activation $\mathbf{H}_i$ in layer $i$ is obtained recursively as an affine map of the activation in the previous layer, composed with some activation function $\phi_i$:
$$
\mathbf{H}_i = \phi_i \left( \mathbf{W}_{i-1,i} \mathbf{H}_{i-1} + \mathbf{b}_i \right), \quad i=1,\ldots,d,
$$
where $d$ is the number of layers of the NN, also called \emph{depth}, $\mathbf{W}_{i-1,i}$ the weight matrix, and $\mathbf{b}_i$ the bias vector. $\mathbf{H}_0$ is given by the $n$-dimensional input $\mathbf{x}_i \in \R^n$, the output of the NN is $\mathbf{H}_d$, and all layers between the input and output are called hidden layers. Each layer has a certain width $w_i$, so the activation $\mathbf{H}_i$ in that layer is a vector of size $w_i$, to which the activation function is applied componentwise. The set of all weight matrices and bias vectors of the network defines the trainable parameters $\vtheta$ of the NN, which have to be optimized through a training procedure on reference data.

NNs of this elementary type have been studied extensively, resulting in a family of \emph{universal approximation theorems}, which discuss the conditions a NN has to satisfy to reproduce a given function \cite{cybenko1989approximation,hornik1990universal,hornik1991approximation}. One of these theorems states that any continuous function $f$ on a compact subset of $\R^n$ can be approximated with arbitrary accuracy by an FCFNN of depth 2 (i.e., with one hidden layer) and with unlimited width and non-polynomial activation function \cite{leshno1993multilayer}. In other words, this guarantees for any "real world function" $f$ that there exists a sequence of depth-two FCFNNs, of whatever setup but appropriate activation function, that uniformly converges to $f$.

\begin{figure}[h!]
	\centering
	\begin{minipage}{0.45\textwidth}
		\includegraphics[width=0.80\textwidth]{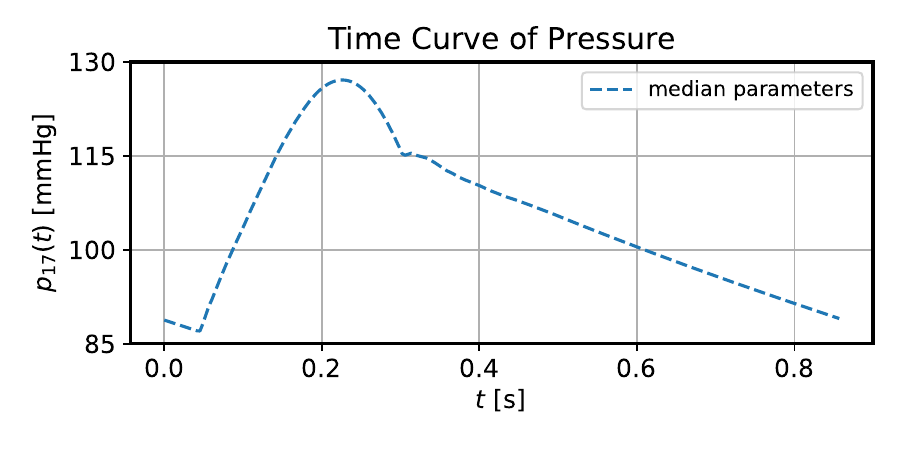}
	\end{minipage}
	\begin{minipage}{0.45\textwidth}
		\includegraphics[width=0.80\textwidth]{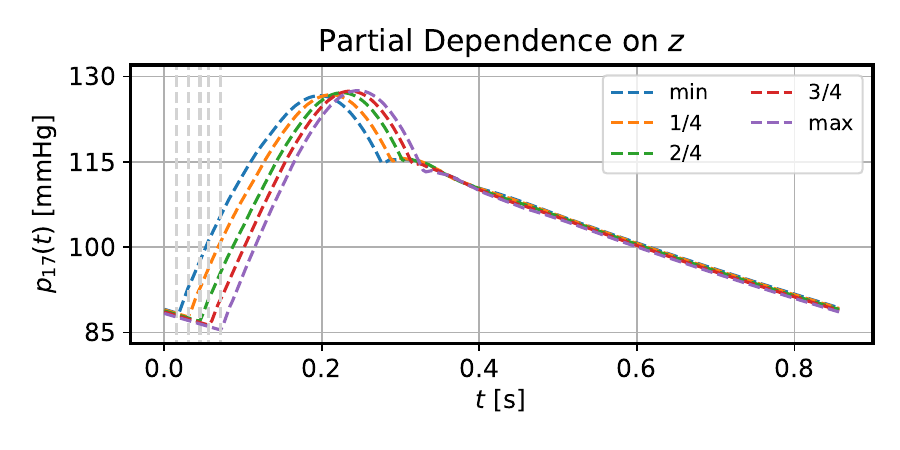}
	\end{minipage}
	\begin{minipage}{0.45\textwidth}
		\includegraphics[width=0.80\textwidth]{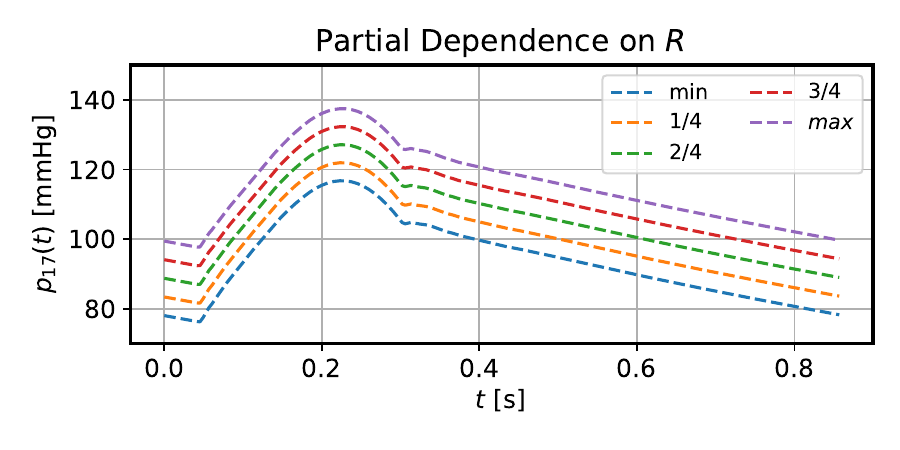}
	\end{minipage}
	\begin{minipage}{0.45\textwidth}
		\includegraphics[width=0.80\textwidth]{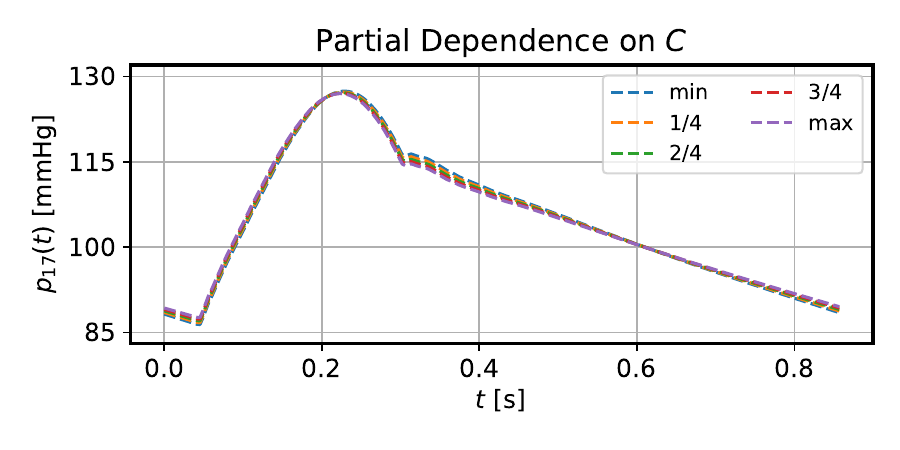}
	\end{minipage}
	\caption{\label{fig:partial_dependencies_p} Characteristic pulse wave and modulation by position, resistance and compliance. Upper left: generic form of a simulated blood pressure pulse wave at median values of $z$, $R$ and $C$ in the simulated range. Upper right, lower left, lower right: modulation by $z$, $R$, and $C$, respectively. Here, the varying parameter covers the respective simulated range in steps of $1/4$, while the other parameters are set to the medians of their ranges. For $z$, dashed lines indicate the phase propagation of the diastolic pressure.}
\end{figure}

\begin{figure}[h!]
	\centering
	\includegraphics[width=0.4\textwidth]{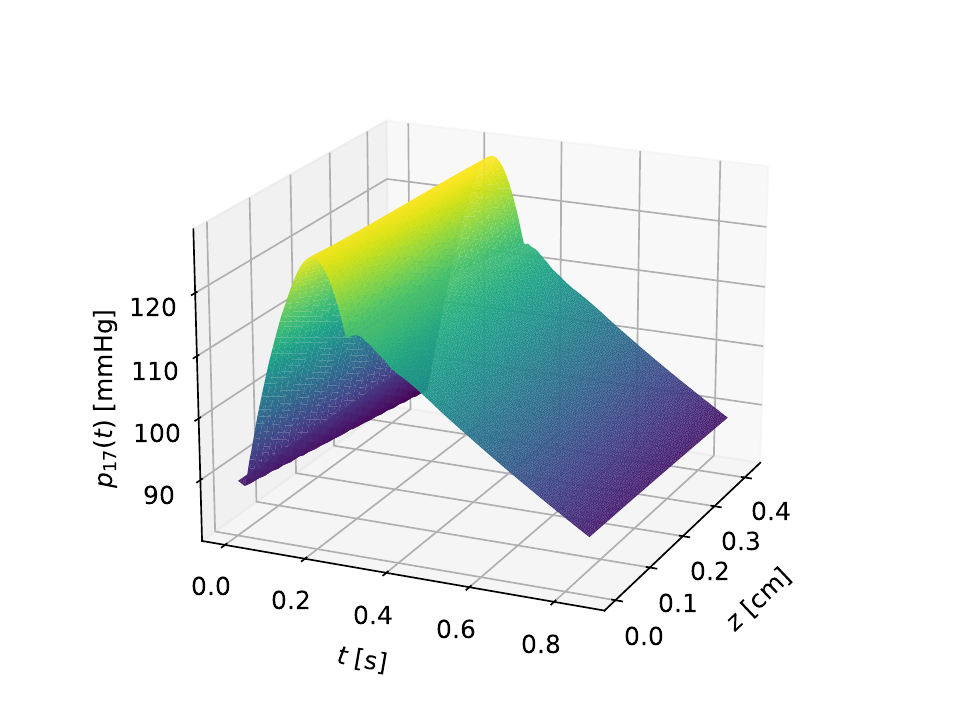}
	\includegraphics[width=0.4\textwidth]{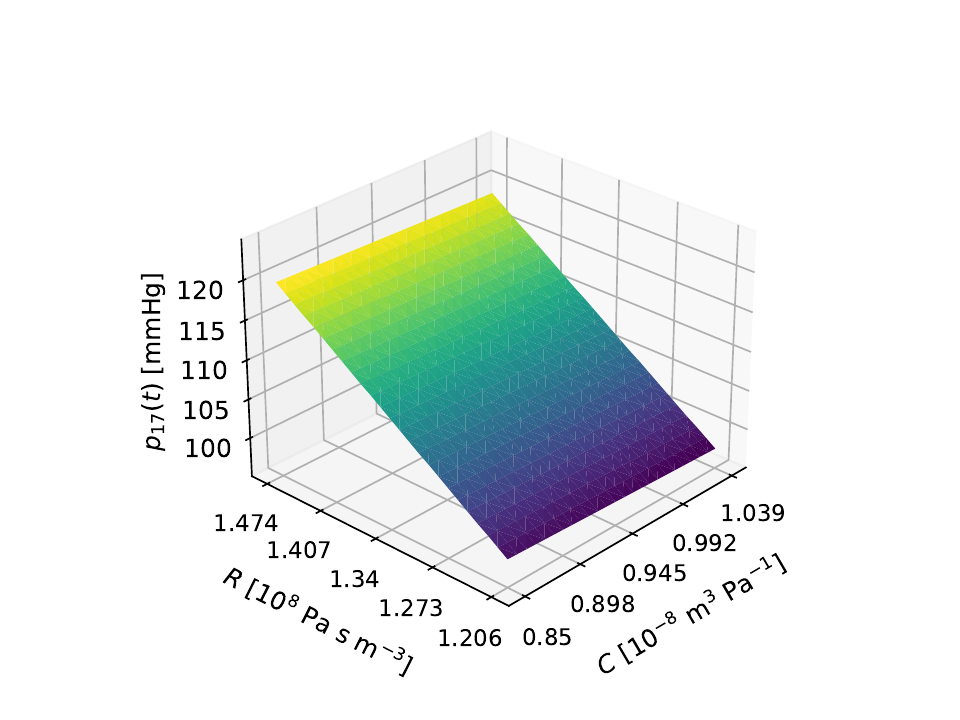}\\
	\includegraphics[width=0.4\textwidth]{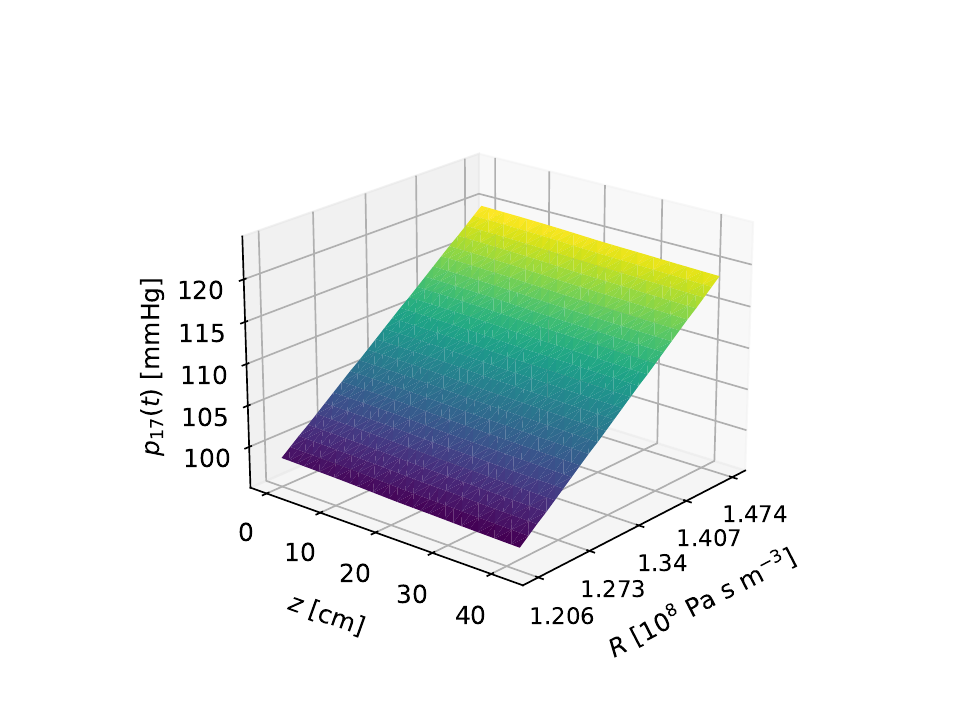}
	\includegraphics[width=0.4\textwidth]{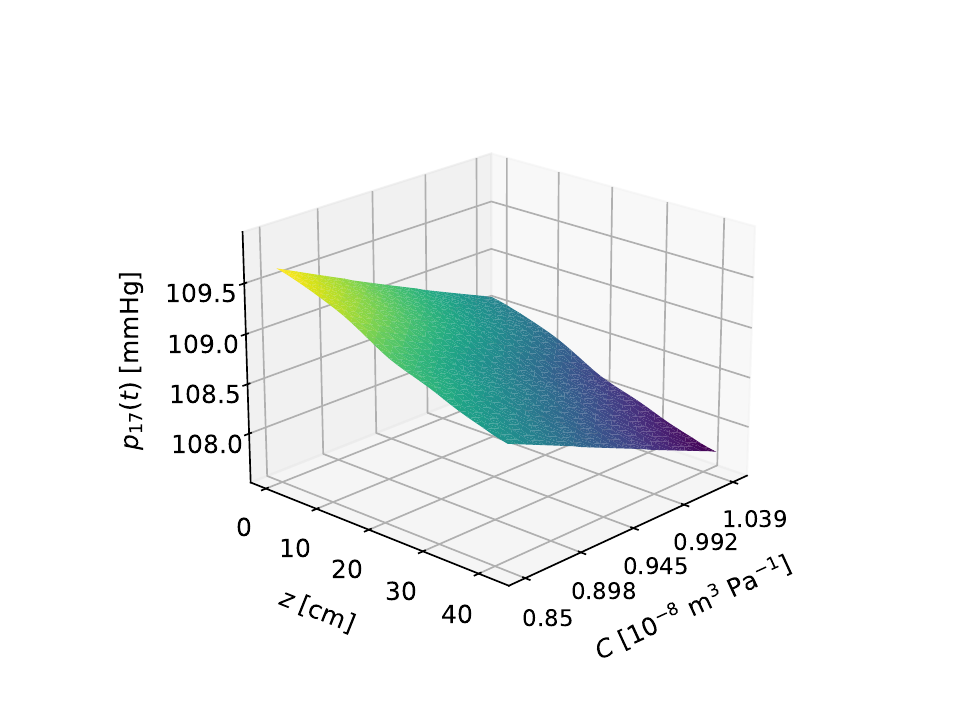}
	\caption{\label{fig:continuity_target} Partial dependency of the simulated pressure on variable pairs. The plots show $p_{17}(z,t,R,C)$ as a function of $t$ and $z$ (top left), of $R$ and $C$ (top right), of $z$ and $R$ (bottom left), and of $z$ and $C$ (bottom right). The respective other two variables are kept fixed at their medians in the simulated range.}
\end{figure}

We briefly justify that this theorem can be applied to the blood pressure curves $p_{17}(z,t,R,C)$ produced by the coupled 1D-0D model of Section \ref{sec:bloodflow}. For this purpose, we provide arguments for the statement that $p_{17}(z,t,R,C)$ is a continuous function, as it is required by the universal approximation theorem. Graphically, the shape of $p_{17}$, as shown in Figs. \ref{fig:partial_dependencies_p} and \ref{fig:continuity_target}, appears essentially smooth and differentiable in the variable $t$. The only exceptions are at the start of the systole (the minimum) and the so-called dicrotic notch at the end of the systole. The same is observed for the variable $z$, i.e., that $p_{17}$ is a smooth and differentiable function with respect to $z$ apart from these kinks. Regarding the remaining two variables, the two WK parameters $R$ and $C$, $p_{17}$ overall even appears to be almost linear. In a more rigorous way, \cite{vcanic2003mathematical} proved that the solution of the blood flow model in Section \ref{sec:bloodflow} is, in fact, continuous in $z$ and $t$ if continuous boundary conditions are imposed. It can be shown for the solution of the 1D model that shocks, breaking the continuity, occur only at distances in the order of several meters \cite{vcanic2003mathematical}, i.e., far beyond the vessel length. Moreover, in \cite[Theorem 3.1]{fernandez2005analysis} the existence of strong (continuous differentiable) solutions for the 1D-0D model is shown. In the referenced work, the 1D-0D coupled problem is formulated as a hyperbolic system of PDEs, where one boundary condition depends on an ODE derived from the 0D WK model. The right-hand side of this ODE depends continuously on the WK parameters, with the consequence that the boundary data and the 1D solution are continuous in the WK parameters. The assumption that $p_{17}$ is continuous in the WK parameters is supported by a Monte Carlo sensitivity analysis of a 1D-OD model for the main arteries in the arm \cite{leguy2011global}. This study demonstrates that the characteristic shape of a pulse wave is only affected by the main model parameters and provides evidence that the resistance and compliance predominantly influence the main features of a pulse wave in a monotonic way. Lastly, the relation between the individual WK parameters of the outlets and the global WK parameters $R$ and $C$, given by \ref{eq:def_Rtot_Ctot}, is continuous. Thus, it is reasonable to assume that $p_{17}$ depends continuously on the inputs $z$, $t$, $R$ and $C$, and that the universal approximation theorem in the version of \cite{leshno1993multilayer} can be applied.

Practically, of course, the full convergence to the target function is not tractable, since its exact functional form is unknown and only accessible at a finite number of supporting points. This puts a principal bar on the approximative quality: a NN with a neuron number in the order of the samples can reproduce these training points perfectly, but in the lack of additional data, one cannot state its accuracy in the interpolating and extrapolating regions, which means that the model might be overfitted. The standard strategy of machine learning to find the optimal model generalizability consists of regularizing the model in some manner to achieve the best trade-off between the model bias and variance. For this work, this means to identify an optimal configuration of the NN widths, depth and activation functions by monitoring the test and validation errors.


Recent research in the field of vascular flow simulations has gone far beyond simple fully connected feedforward NNs \cite{nolte2022inverse}. Here, \emph{physics-informed neural networks} (PINN) are prominent \cite{kissas2020machine}, which enhance the data-driven training through regularization that enforces consistency with the underlying model equations. In fact, this idea has been successfully applied to a vascular system with a single bifurcation, modeled by the 1D-OD blood flow model, in \cite{kissas2020machine}. The regularization uses the residuals of the partial differential equations \ref{eq:1D_PDEs}. However, the resulting FCFNN comprises seven layers of width $100$ (\cite[Section 3.1.3]{kissas2020machine}). This results in a huge number of parameters that need to be trained. This illustrates a general disadvantage of PINNs: their high computational cost, caused by the required large network size and the repeated evaluation of the residuals. Moreover, their implementation and hyperparameter setup are challenging and lead to a significant methodological overhead.

Despite recent works improving the flexibility, efficiency, and stability of PINNs\cite{lawal2022physics}, there is, to our knowledge, no drastically simplified and accelerated modification of PINNs for modeling small cardiovascular systems. In view of these drawbacks, we use a straightforward, purely data-driven FCFFN, ideally of small size. A surrogate model of this type may maximally speed up the calibration procedure, as required for real-world applications. Thus, the paradigm of our work is to provide the model with all necessary information on the arterial network solely in the form of data and, from this, distill a minimal-sized FCFNN. The training data will be produced by solving the model equations  \eqref{eq:1D_PDEs} and \eqref{eq:PressureAreaRelation}. This means that regularizing the NN using the residuals of these PDEs would be redundant, as long as the data is sufficiently resolved. In this sense, the effort that the PINN approach spends on regularization by the physical equations is shifted to generating reference data from the physical equations. Naturally, this implies that data quality is critical for our approach, but it keeps the model structure significantly simpler.

\subsection{Reference data} 
\label{subsec:data}
As pointed out above, our method requires a high-quality and representative data set of the 1D-0D blood flow model. We have generated such a data set anew by applying the numerical solution techniques of \ref{subsec:solver} to solve the 1D-OD flow model and report the data in Vessel $17$. For this single artery, the variable space is four-dimensional, given by the time $t$, the position $z$ and the two global WK parameters. In this case, it is computationally feasible to simulate the system on dense sampling meshes (also see \ref{subsec:solver} concerning NMC).

In detail, we consider a full cardiac cycle of constant period $T=0.857\;\unit{s}$, and compute this on 228 sampling steps. The position along the artery of length $42.2\;\unit{cm}$ is sampled on 31 points. For the two global WK parameters, we employ a $25 \times 25$ grid. The resistance parameter is varied around $1.34$ in the range $[1.206,1.474]$ (in $10^8\;\unit{Pa\;s}\;\unit{m}^{-3}$) and the compliance around $0.945 \cdot 10^{-8}$ in the range $[0.850,1.039]$ (in $10^{-8}\;\unit{m}^{3}\;\unit{Pa}^{-1}$). The values at the centers of both ranges are standard values for the total resistance and compliance of the systemic arterial system in normal conditions \cite[Section 2.2]{alastruey2007modelling}. Varying the WK parameters within these ranges allows us to cover a physiological range for the blood pressure curves in the network.

The simulation actually outputs the cross sectional area and the volume flow rate, and the pressure is calculated from the former via the relation \eqref{eq:PressureAreaRelation}, where the constants cross-sectional area at rest $A_0$ and the characteristic pressure parameter $G_0$ take the values reported in Table \ref{tab:DataNetwork}. For the first pair of WK parameters $\left(R,C\right)$, the initial values $A=0$, $p=0$ and $Q=0$ are used in each vessel, and the system is then simulated over ten cardiac cycles until the patterns for $A$, $p$, and $Q$ become periodic. For all subsequent $\left(R,C\right)$ pairs, the system is initialized using the state at the last time step of the previous $\left(R,C\right)$ pair. In this way, we need to simulate only approximately five cardiac cycles to reach stable behavior. For each parameter pair, we record only the final simulated period and include this in the training data. It is also important to note that we compute the pulse waves along the entire arterial length simultaneously, i.e., at all grid points of $z$, for each pair of WK parameters, which captures the pulse propagation illustrated in Figure \ref{fig:phase_shift}. Thus, the reference data set in comprises $31 \times 25 \times 25$ pulse waves of 228 time steps each, which is 4,417,500 single pressure records. The details of the grid are also summarized in Table \ref{tab:simulation_details}.
\begin{table}[h!]
\begin{tabular}{|c|r|c|r|r|r|}
\hline 
 & unit & grid points & min & max & mesh width\tabularnewline
\hline 
$t$ & s & 228 & 0 & 0.857 & 0.00375\tabularnewline
\hline 
$z$ & cm & 31 & 0 & 42.2 & 1.4\tabularnewline
\hline 
$R$ & $10^8\;\unit{Pa}\;\unit{s}\;\unit{m}^{-3}$ & 25 & 1.206 & 1.474 & 1.1\tabularnewline
\hline 
$C$ & $10^{-8}\;\unit{m}^{3}\;\unit{Pa}^{-1}$ & 25 & 0.850 & 1.039 & 0.008\tabularnewline
\hline 
\end{tabular}
\caption{Number of sampling points and ranges of the different input variables in the reference data.}
\label{tab:simulation_details}
\end{table}

\subsection{General setup of the neural network}
As motivated in Section \ref{sec:justification}, the surrogate model is constructed as a multilayer FCFNN with a structure shown in Figure \ref{fig:NN_enhanced}. The input is given by the four variables $(z,t,R,C)$, which is then processed in $d-1$ hidden layers of equal width $w$ and specific activation functions $\sigma_i$. Importantly, we allow different activation functions across the layers, while keeping them the same within a given layer. The last layer, with index $d$, is constructed as a single linear unit, which produces the output, i.e., the prediction $\hat{p}_{17}$. Thus, a specific architecture is fully described by the depth $d$, the width $w$, and the type of activation functions of the hidden layers. It may seem appealing to construct a more complex NN, that implements the hierarchy of the variables, e.g., by first learning a generic time-curve, and then learning how to modulate this time-curve with respect to the other variables. However, some first tests with this kind of NN did not pay off in accuracy compared to the simple FCFNN structure.

The training procedure is based on minimizing the mean square error loss
$$
\mathcal{L}(\vtheta) = \frac{1}{n} \sum_{i=1}^{n} \Big( f_{\vtheta}(z_i,t_i,R_i,C_i) - p_{17}(t_i,z_i,R_i,C_i) \Big)^2,
$$
measuring the mismatch between the NN prediction $f_{\vtheta}$ at the trainable parameters $\vtheta$ and the corresponding simulated reference values $p_{17}$ at $n$ samples. For the parameter update, we employ the Adam optimizer \cite{kingma2014adam}, a stochastic gradient descent method well suited for training on large data sets. Generally, we split the optimization into two phases of a certain number of epochs, where the first uses a larger training rate $\eta_1$ and the second a smaller one that is denoted by $\eta_2$. All epochs process the training data in mini-batches of a certain size and consist of a full cycle over all mini-batches.

The total training data are split into a training set of size 80$\%$ and the remaining 20 $\%$ are used for validation, in particular to test for overfitting, and model selection. Here, the random assignment is done once and then used for all studied cases. We did not use a separate hold-out test set, as is usually recommended. In view of the highly homogeneous reference data and -- as will be apparent below -- the high agreement of the test and validation errors, this is tolerable. Prior to the training, the network weights are initialized by Glorot's scheme\cite{glorot2010understanding} and the biases are set to zero. The NN is implemented using the TensorFlow Keras framework (version 3.4.1), and the computations could be run on standard machines without GPU usage.

As an important preprocessing step, all input variables (and also the target) are standardized to have zero mean and variance one. This practice typically improves the convergence speed and stability of the training, since it brings all inputs to approximately the same order of magnitude \cite{glorot2010understanding}. In \cite{kissas2020machine}, it has been demonstrated that this preprocessing step is beneficial at learning blood flow data. Note that we determine the mean values and standard deviations for the standardization on the total data set, including the validation part, and use these in the transformation of any input. This is valid, i.e., it does not leverage information between training and testing, since the standardization effectively just defines a different scale for each physical dimension. Since the NN implicitly depends on this scale, the transformation cannot be changed. The output of the NN, the blood pressure, is also produced on a standardized scale and then, if required for evaluation, transformed back to physical units (mmHg). To quantify the prediction accuracy, we report the mean absolute error (MAE) between the NN outputs and the corresponding reference values. While MAE and root mean square error show generally highly consistent trends within the individual tests, we prefer MAE due to its greater robustness to outliers in the prediction. This property is favorable for our analysis, where we compare different subsets of the training data where the error distributions might vary. So, although the NN is trained on mean square error, the tests below use the MAE, due to its robustness and better interpretability as a metric with the same dimension as the target quantity.

The following sections present a series of numerical tests to identify an optimal configuration of the NN surrogate, concerning the network depth, width, and employed activation functions. To accelerate the search of these model hyperparameters at this intermediary stage, the optimization procedure is changed to first training the NN on 10 reproducible random subsets, each containing only $5\%$ of the training samples and then fine-tune the NN on the total training set (first parts: batch size 128, each 100 epochs, second part: batch size 512, again 100 epochs). We determined by an additional test that this procedure yields MAE values that are stable within $\pm 0.005 \, \unit{mmHg}$, i.e., that the third decimal digit is not fully reliable. We remark that using a fixed number of epochs while varying network size may disadvantage larger networks that converge more slowly. However, this aligns with our goal of identifying a parametrically efficient model.
Throughout the study, we use an MAE $0.020 \, \unit{mmHg}$ (MAE) as a performance benchmark. This value was deliberately stringent, since the model is expected to generally perform well during the diastolic phase, where the pulse wave is flat and almost linear. Even if there are significant deviations apart from that, the diastolic phase is dominating since it covers most of the time points. Thus, we argue that achieving or surpassing this benchmark would indicate that the NN surrogate is accurate enough to safely replace the actual simulation in the calibration.

\subsection{Changing the activation function}
To incorporate the shape of the pulse wave, which exhibits some smooth parts and two pronounced kinks, we investigated the optimal configuration of the activation functions. Here, we argue that the rectified linear unit (ReLU, \cite{nair2010rectified}), a piece-wise linear function, is well-suited to capture the linear segments and the two kinks (the first at the diastole and the second shortly after the systole, the so-called dicrotic notch). Conversely, the hyperbolic tangent ($\tanh$) is ideal for modeling the non-linear parts of the pulse wave, in particular the systolic upstroke and decline. Our hypothesis is that a combination of these two activation functions would best capture these characteristics.

\begin{figure}[h!]
\begin{minipage}{0.45\textwidth}
\begin{center}
(a) ReLU $\rightarrow$ ReLU, $1.248 \, \unit{mmHg}$ (MAE)
\includegraphics[scale=0.4]{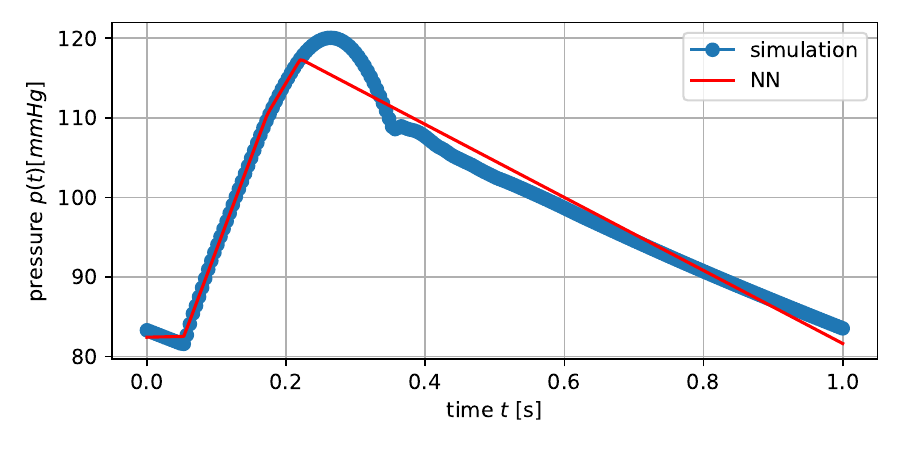}
\end{center}
\end{minipage}
\begin{minipage}{0.45\textwidth}
\begin{center}
(b) $\tanh$ $\rightarrow$ $\tanh$, $0.524 \, \unit{mmHg}$ (MAE)
\includegraphics[scale=0.4]{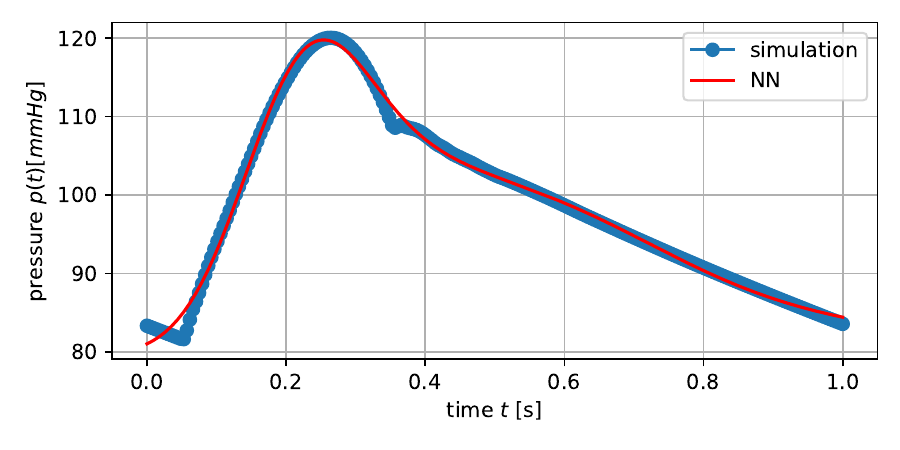}
\end{center}
\end{minipage}
\begin{minipage}{0.45\textwidth}
\begin{center}
(c) $\tanh$ $\rightarrow$ ReLU, $1.040 \, \unit{mmHg}$ (MAE)
\includegraphics[scale=0.4]{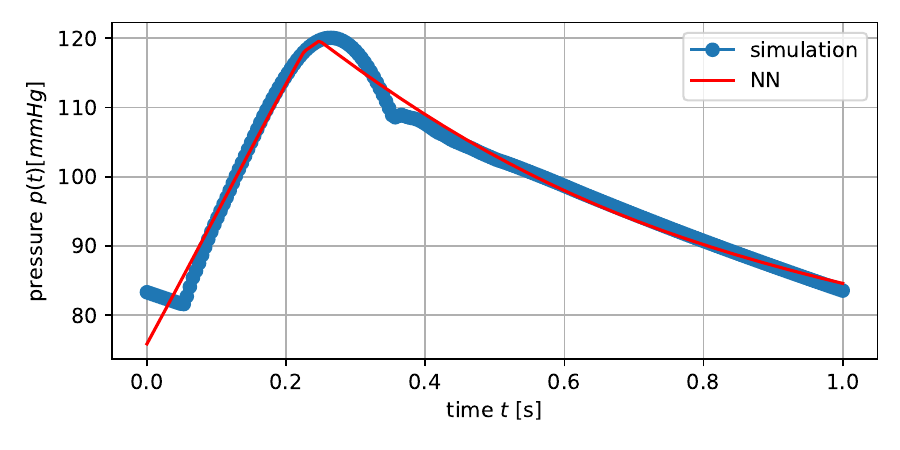}
\end{center}
\end{minipage}
\begin{minipage}{0.45\textwidth}
\begin{center}
(d) ReLU $\rightarrow$ $\tanh$, $0.343 \, \unit{mmHg}$ (MAE)
\includegraphics[scale=0.4]{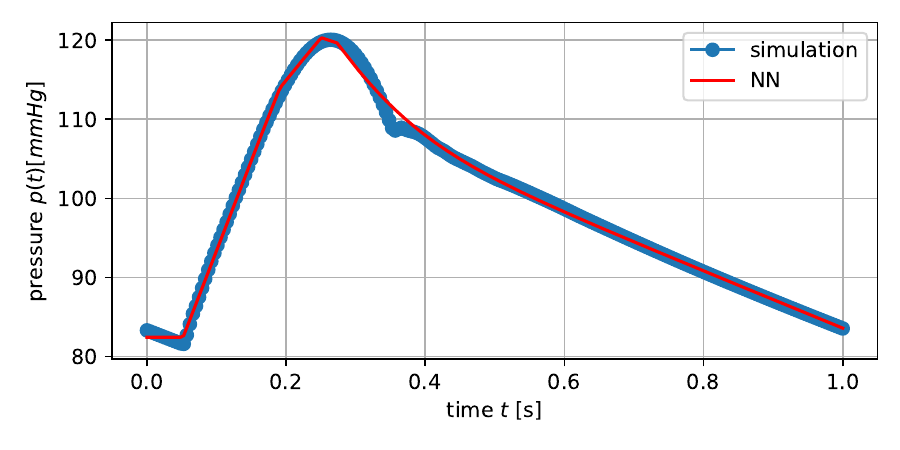}
\end{center}
\end{minipage}
\caption{\label{fig:choice_of_act} Modeling a pressure curve by different activation function. The pressure $p_(t)$ is learned at $t=l/2$, $R_{1/2}$ and $C_{1/2}$ by a FFNN with two layers of width 6, with pure ReLU (a), pure $\tanh$ activation (b), and with mixed activation with first $\tanh$ (c) and first ReLU. ($\eta=0.001$, batch size 32, 1000 epochs).}
\end{figure}

To underpin this idea, we conducted a numerical experiment using a small NN with two hidden layers of width 6 (and a linear unit) to learn a single pressure curve. This experiment is designed to highlight the differences between different configurations of activation functions. As shown in Figure \ref{fig:choice_of_act}, a pure ReLU network yields an overly rigid piecewise-linear approximation (a), while a pure $\tanh$ network produces a smoother but imprecise prediction, especially at the kinks (b). In contrast, a network with a mixture of activation functions may achieve a better approximation. However, it seems to be important that the ReLU is in the first hidden layer and $\tanh$ in the second (d) while the alternate order is still deficient (c).

To validate these assumptions more systematically, we perform another test on the full dataset at various network depths (2, 3, and 4 hidden layers) at a fixed width of 32 neurons. Our results confirm that a mixture of activation functions generally outperforms pure ReLU or $\tanh$ configurations (see Table \ref{tab:test_activation_functions}). For networks with two hidden layers, the ReLU $\rightarrow$ $\tanh$ configuration is best, just as in the smaller experiment above, for networks with three hidden layers, ReLU $\rightarrow$ ReLU $\rightarrow$ $\tanh$ performed best. In the case of deeper NNs (with four hidden layers), the improvement was less significant due to the large number of trainable parameters, which can obscure the effect of different activation functions.

In total, our findings suggest that placing ReLU activation in earlier layers combined with $\tanh$ in the subsequent layers is beneficial. This might be the case because it can model the discontinuities closer to the time-domain input and the non-linearities closer to the pressure-domain output. Overall, the impact of the activation function configuration was less pronounced in the full-variable-space models compared to the initial numerical experiment, likely due to the larger network size and the effect of the other variables, which act predominantly in a linear way on the pressure (see Figure \ref{fig:continuity_target}).

\begin{table}[htb!]
\begin{minipage}{0.45\textwidth}
\begin{center}
\begin{tabular}{|c|c|c|c|}
\hline 
\multicolumn{2}{|c|}{layer} & \multicolumn{2}{c|}{}\tabularnewline
\hline 
1 & 2 & \multicolumn{2}{c|}{MAE}\tabularnewline
\hline 
\multicolumn{2}{|c|}{activation} & train. & val.\tabularnewline
\hline 
tanh & tanh & 0.027 & 0.027 \tabularnewline
\hline 
tanh & ReLU & 0.035 & 0.035 \tabularnewline
\hline 
\textbf{ReLU} & \textbf{tanh} & \textbf{0.023} & \textbf{0.023} \tabularnewline
\hline 
ReLU & ReLU & 0.036 & 0.036 \tabularnewline
\hline 
\end{tabular}
\end{center}
\end{minipage}
\begin{minipage}{0.45\textwidth}
\begin{center}
\begin{tabular}{|c|c|c|c|c|}
\hline 
\multicolumn{3}{|c|}{layer} & \multicolumn{2}{c|}{}\tabularnewline
\hline 
1 & 2 & 3 & \multicolumn{2}{c|}{MAE}\tabularnewline
\hline 
\multicolumn{3}{|c|}{activation} & train. & val.\tabularnewline
\hline 
tanh & tanh & tanh & 0.036 & 0.036 \tabularnewline
\hline
tanh & tanh & ReLU & 0.032 & 0.032 \tabularnewline
\hline
tanh & ReLU & tanh & 0.020 & 0.020 \tabularnewline
\hline
tanh & ReLU & ReLU & 0.019 & 0.019 \tabularnewline
\hline
ReLU & tanh & tanh & 0.021 & 0.021 \tabularnewline
\hline
ReLU & tanh & ReLU & 0.021 & 0.021 \tabularnewline
\hline
\textbf{ReLU} & \textbf{ReLU} & \textbf{tanh} & \textbf{0.016} & \textbf{0.016} \tabularnewline
\hline
ReLU & ReLU & ReLU & 0.020 & 0.020 \tabularnewline
\hline 
\end{tabular}
\end{center}
\end{minipage}
\caption{Training and validation MAE at different configurations of the activation functions in the inner layers. Left: depth-2 cases, right: depth-3 cases.}
\label{tab:test_activation_functions}
\end{table}

\subsection{Confining the topology}
Next, we investigate the network topology, where we consider only networks of equal widths in all hidden layers. This reduces the search space, and, based on prior experience, this restriction does not limit the performance of the model for regression tasks. We specifically avoided a bottleneck architecture because the smaller widths close to the output could promote an oversimplified and underfitted approximation of the fine details of the pulse wave. To determine the effect of the network topology, we tested setups with two, three and four hidden layers (excluding the last linear unit) where we also varied the widths of the hidden layers from 8 to 128 (in powers of 2). Based on the previous analysis, we test the configurations ReLU-$\tanh$, ReLU-ReLU-$\tanh$ and ReLU-ReLU-$\tanh$-$\tanh$ of the activation functions. The resulting training and validation errors are listed in Table \ref{tab:test_architecture}.

There are two main observations in the results: (i) The models are not overfitting across all tested setups, even for the larger networks. This follows from the training and validation errors, which are consistent. This is likely due to the fine-sampled and homogeneous nature of our dataset, where the larger training and the smaller validation part typically represent the total data similarly well. (ii) All setups achieve a satisfactory performance. Even the smallest ones surpass an MAE of $0.10 \, \unit{mmHg}$. This might be related to the simplicity of the dominating structure of the underlying problem, such that a relatively small network can already capture the essential features of the problem. Going into more detail about the results, we see that increasing either the depth or the width improves the predictive performance. Here, it is difficult to compare the changes in depth or in width since they do not increase the number of trainable parameters in the same way. However, it seems that the improvements are mainly driven by increasing the number of parameters, rather than their actual distribution in depth or width of the network.

The gain in predictive performance due to increased network size reaches a plateau of ca. $0.01 \, \unit{mmHg}$. Here, the optimum is not attained by the largest setup (4 hidden layers of width 128), but by a setup with 3 hidden layers of width 128. However, the first, i.e., the smallest, architecture with a performance below $0.020 \, \unit{mmHg}$ -- the previously defined as a benchmark -- is given a 3 layer topology with width 32. We selected this architecture as our optimal model since it ideally combines high efficiency and accuracy. It is remarkable that this configuration possesses only 2,305 parameters, meaning that it compresses the information in the data to less than $0.07\%$ of the number of training samples.

\begin{table}[htb!]
\begin{minipage}{0.3\textwidth}
\begin{center}
\begin{tabular}{|c|c|c|}
\hline 
\multicolumn{3}{|c|}{ReLU-$\tanh$}\tabularnewline
\hline 
 & \multicolumn{2}{c|}{MAE}\tabularnewline
\hline 
width & train. & val.\tabularnewline
\hline 
8 & 0.070 & 0.070 \tabularnewline
\hline
16 & 0.031 & 0.031 \tabularnewline
\hline
32 & 0.023 & 0.023 \tabularnewline
\hline
64 & 0.016 & 0.016 \tabularnewline
\hline
128 & 0.017 & 0.017 \tabularnewline
\hline 
\end{tabular}
\end{center}
\end{minipage}
\begin{minipage}{0.3\textwidth}
\begin{center}
\begin{tabular}{|c|c|c|}
\hline 
\multicolumn{3}{|c|}{ReLU-ReLU-$\tanh$}\tabularnewline
\hline 
 & \multicolumn{2}{c|}{MAE}\tabularnewline
\hline 
width & train. & val.\tabularnewline
\hline 
8 & 0.046 & 0.046 \tabularnewline
\hline
16 & 0.035 & 0.035 \tabularnewline
\hline
\textbf{32} & \textbf{0.016} & \textbf{0.016} \tabularnewline
\hline
64 & 0.018 & 0.018 \tabularnewline
\hline
128 & 0.012 & 0.012 \tabularnewline
\hline 
\end{tabular}
\end{center}
\end{minipage}
\begin{minipage}{0.3\textwidth}
\begin{center}
\begin{tabular}{|c|c|c|}
\hline 
\multicolumn{3}{|c|}{ReLU-ReLU-$\tanh$-$\tanh$}\tabularnewline
\hline 
 & \multicolumn{2}{c|}{MAE}\tabularnewline
\hline 
width  & train.  & val.\tabularnewline
\hline 
8  & 0.042 & 0.042\tabularnewline
\hline 
16  & 0.024 & 0.024\tabularnewline
\hline 
32  & 0.019 & 0.019\tabularnewline
\hline 
64  & 0.013 & 0.013\tabularnewline
\hline 
128  & 0.014 & 0.014\tabularnewline
\hline 
\end{tabular}
\end{center}
\end{minipage}
\caption{Variation of the neural network architecture. Reported are the training and validation MAEs for NNs with 2, 3, and 4 hidden layers in the configuration ReLU-$\tanh$ (left), ReLU-ReLU-$\tanh$ (center), and ReLU-ReLU-$\tanh$-$\tanh$ (right). The widths of these layers are varied between 8 and 128, as stated.}
\label{tab:test_architecture}
\end{table}

\subsection{Performance of the optimal set-up}
In the previous parts, it has been worked out that an appropriate variant of the NN has 3 hidden layers of width 32 with the activation functions in the configuration ReLU-ReLU-$\tanh$. To ensure full convergence of the neural network, it was retrained for 3,000 epochs at learning rate $\eta=0.001$ and batch size 512. Since this model serves as the basis for most studies of the subsequent calibration below, it is important to assess its performance.

Table \ref{tab:surrogate_performance} reports its training and validation MAE, and, additionally, the worst-case MAE among all predicted pressure curves, i.e., of all tested parameter triplets $(z,R,c)$. Compared to the accelerated but rougher training procedure from above, the extended training dropped the errors moderately. The final surrogate achieves an MAE of $0.011 \, \unit{mmHg}$ (training) and $0.012 \, \unit{mmHg}$ (validation), comfortably meeting our accuracy demands. Even the worst-case MAE in a full pulse wave is only $0.024 \, \unit{mmHg}$, only slightly exceeding the $0.020 \, \unit{mmHg}$ benchmark.

\begin{figure}[h!]
\begin{center}
\includegraphics[scale=0.6]{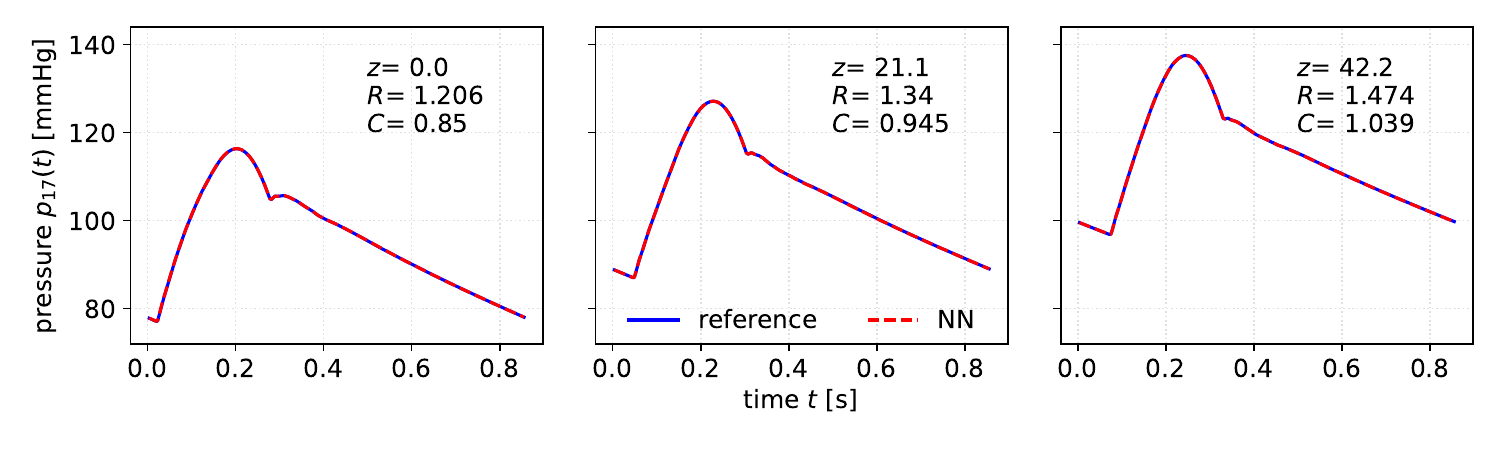}
\end{center}
\begin{center}
\includegraphics[scale=0.6]{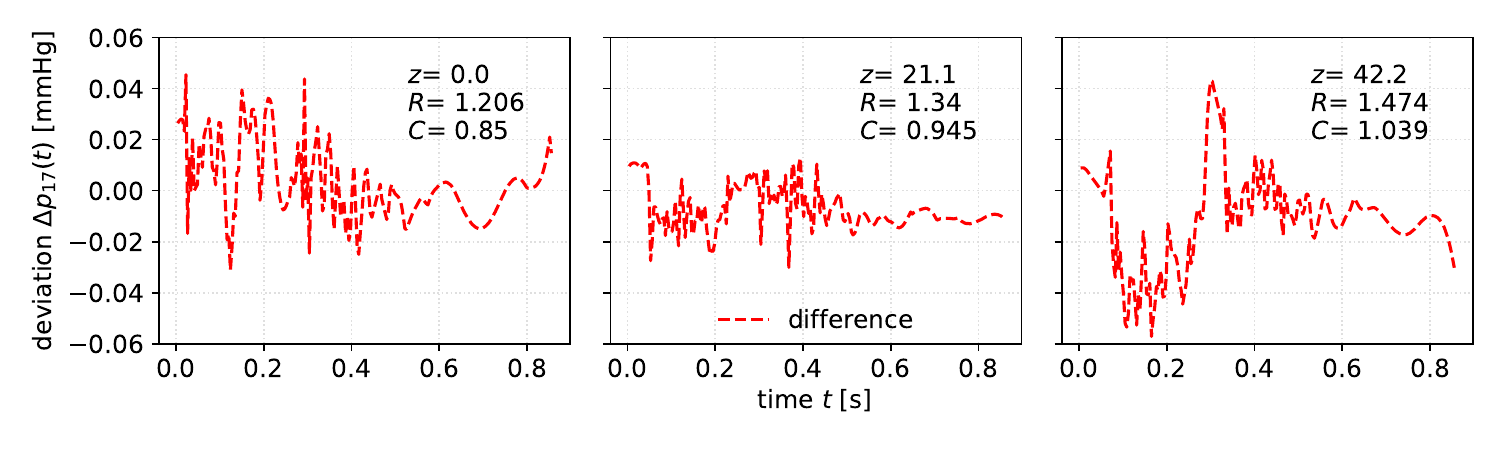}
\end{center}
\caption{Comparison of the final model's prediction with the reference data. Upper panel: predicted (dashed red) and reference pulse wave, lower panel: pointwise deviation of the curves. The two left plots refer to the respective minimum values of the parameters $z$, $R$ and $C$, the two middle to the median values, and the two right to the maximum values. Units: $z$ in $\unit{cm}$, $R$ in $\uR$, $C$ in $\uC$.}
\label{fig:surrogate_performance}
\end{figure}

Figure \ref{fig:surrogate_performance} illustrates the model accuracy by comparing the prediction with the target data at the combined minimum, median and maximum of the considered parameter ranges. Here, the upper panel shows the predicted and target pulse waves, which are visually indistinguishable -- the model reproduces all characteristics of the pulse waves with high fidelity, in particular the kinks at the diastole and the dicrotic notches. The lower panel presents the pointwise differences between prediction and target, which are patternless, except for the diastolic phase, where the differences are more stable, as expected for the close-to-linear pressure decline in that phase. More importantly, the deviations remain in a very narrow range with a span of about $0.1 \, \unit{mmHg}$, which confirms the high agreement of model and target.

In view of the model efficiency, evaluating a pulse wave (at 228 time steps) requires about $1.0 \, \unit{ms}$ if executed on a standard notebook without GPU usage and in graph mode. Generating the same pulse wave with the full 1D-0D model would take about 10 seconds of wall time. Thus, using the surrogate has the potential to speed up the calibration by a factor of approximately $10^{-4}$. Since it is, at the same time, highly accurate, we expect it to reliably replace the explicit simulation in the WK parameter estimation.

\begin{table}{htb!}
\begin{tabular}{|l|c|c||c|c|}
\hline 
 & train. & val. & min. & max.\tabularnewline
\hline
MAE [mmHg] & 0.011 & 0.012 & 0.005 & 0.020\tabularnewline
\hline
\end{tabular}
\caption{Performance of the parametrized network architecture. Listed are the training and validation error and the best and the worst prediction of a single pressure curve at fixed $z$, $R$ and $C$, in terms of the mean absolute error (MAE).}
\label{tab:surrogate_performance}
\end{table}

\subsection{Sensitivity to the training set size}
\label{subsec:NN_less_data}
\begin{figure}[h!]
\begin{center}
\includegraphics[scale=0.5]{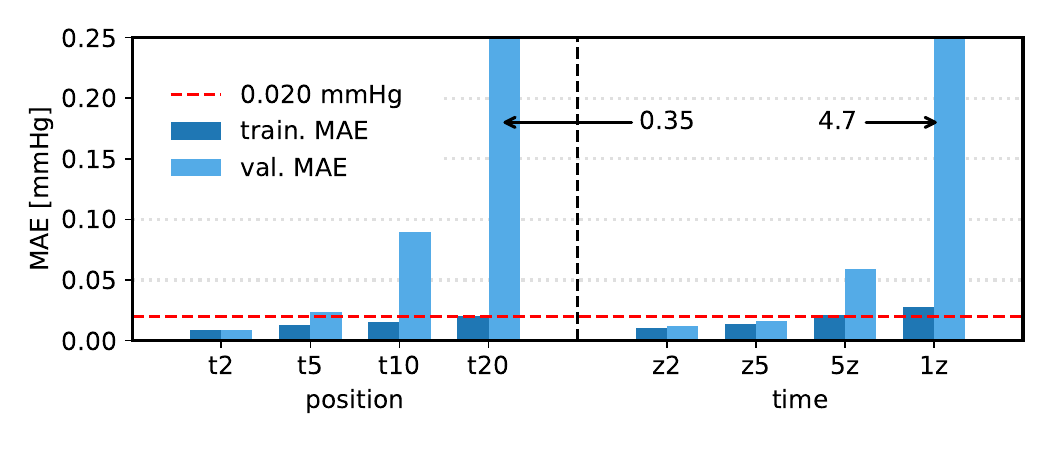}
\end{center}
\begin{center}
\includegraphics[scale=0.5]{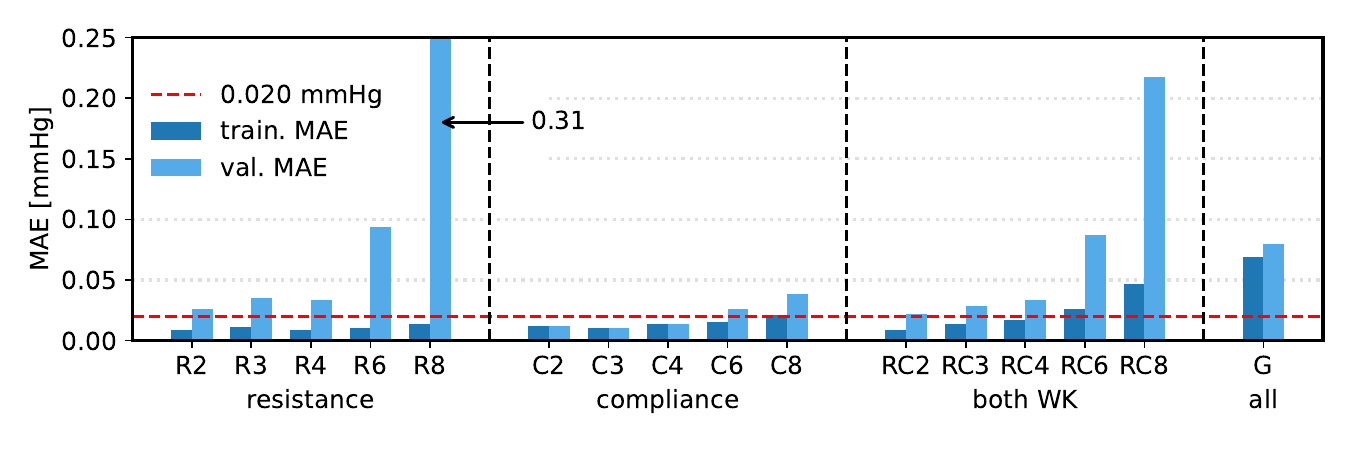}
\end{center}
\caption{Predictive performance of the NN surrogate at reduced training data. The bars show the MAE on the referring training and validation parts of the tested reduction cases (for the explanation, see the plain text). The red line indicates the desired accuracy benchmark of $0.020 \, \unit{mmHg}$.}
\label{fig:thinning_NN}
\end{figure}

Our purely data-driven approach requires a fully representative reference data set. It is therefore important to understand how much the reference data can be reduced -- along time $t$, position $z$, and the WK parameters $R$ and $C$ -- without substantially degrading the NN's performance. This also informs us how dense new data should be simulated. To analyze this, we retrained the optimal setup (ReLU-ReLU-$\tanh$, width 32) on increasingly thinned versions of the complete dataset (see Section \ref{subsec:data}). Note that we used subsets of the original simulation, only selecting data samples in a regular manner, but not reducing the resolution of the simulations themselves (e.g., by larger time and position increments). This avoids additional simulations, but might underestimate the resulting lowered precision, since truly lower-resolved data would also introduce an error from the coarser solver grids. For the WK parameters $R$ and $C$, this is not an issue since they only act as parameters in the simulation. The structure of the neural network remained the same throughout -- we only adapted its trainable parameters to estimate how a fixed structure responds to reducing the data. If changing to different arterial systems, it could be necessary to modify the topology. For each case, the NN was trained for 1000 epochs on the respective subset of the data and evaluated on the corresponding remainder. This means that the ratio of training and validation data varies, depending on the data thinning (here, using MAE instead of RMSE might be advantageous). Most reductions increase the mesh spacing by a factor (e.g., "t5" denotes that every 5th time step is used). The tested cases are:
\begin{itemize}
\item time grid: t2, t5, t10, t20
\item spatial grid: z2, z5, 5z (only inlet, center, outlet, and two positions between), 1z (center only)
\item Windkessel parameters: R2, R3, R4, R6, R8; C2, C3 C4, C6, C8
\item combined Windkessel parameters: RC2, RC3, RC4, RC6, RC8 
\item global tinning (G): $t$ and $z$ samples five times coarser, $R$ and $C$ four times coarser.
\end{itemize}
Figure \ref{fig:thinning_NN} shows the training and validation MAEs for all variants, along with the desired benchmark of $0.020 \, \unit{mmHg}$. As expected, training errors remain low in most cases -- the model just adapts to fewer data -- and increase only slightly for thinner data sets. Likely, this is because the network becomes over-dimensioned for a small dataset and hence requires more epochs to converge. Validation errors reflect how the model generalizes in the interpolative parts. Coarsening either the time or the spatial grid has only little impact on the validation error if the factors are 5 or less. 
Training at five of the 31 positions (5z) still yields acceptable predictions elsewhere, but training only at the center of the vessel fails. This is because the pulse propagation is present in the reference data, which, however, the NN cannot capture without spatial variation. The resolution of $R$ can be reduced by a factor of 4. For $C$, the overall effect on the errors is smaller, consistent with its weaker influence on the pressure dynamics. Using factors up to 4 basically keeps the accuracy level of the fully-resolved data. Jointly coarsening $R$ and $C$ by a factor of 4 (RC4) similarly remains acceptable, which is unsurprising, since the effects of $R$ are dominating, also if $C$ is thinned.

The global thinning (G) combines the admissible resolutions found so far, which reduced the dataset to just 0.4\% of the original size. While this results in increased errors (MAEs of $0.07 \, \unit{mmHg}$ and $0.08 \, \unit{mmHg}$), the performance remains in an acceptable range. Thus, the NN surrogate can be trained reliably on a small fraction of the original data, which would save a lot of computation costs when simulating new reference sets. These results also clarify the limits of our purely data-driven approach: reducing the sampling density by more than a factor of 4 - 5, the data set does not provide enough information to fully represent the dynamics in the vessel. This motivates introducing some sort of regularization of the NN, e.g. by the governing PDEs as in the PINN approach (see Section \ref{sec:justification}), to successfully train a NN also in case of sparse data.

We also examined how the surrogate behaves in the extrapolating region, i.e. beyond the sampled parameter ranges of the original data. Specifically, we evaluated the NN for values of $R$ and $C$ extending the respective sampled ranges by approximately $\pm 30 \%$ of the lower or upper boundaries (note that extrapolation in $t$ or $z$, i.e. out of the period or the dimension of the vessel, is not meaningful). We found that the predicted pulse wave remain reasonable only in a within a narrow band around the sampled region. If using the full-data version of the NN, this admissible extrapolation margin has the width of ca. 4 - 10 times the original sampling step. Importantly, the overall shape of the predictions does not deteriorate suddenly in the extrapolating regions but gradually gets less reliable. This suggests that the model structure itself keeps suitable but requires for more data to accurately adapt to these regions as well. In this new data, the next data points should be placed at intervals similar to the minimum tolerable resolution found above, i.e. about 4-5 times the original sampling step. Lastly, we point to Sec. \ref{subsec:test_calibration_less_data} that discusses how the use of reduced-data models indirectly affects the calibration.

\subsection{Enhancing the neural network for calibration}
\label{subsec:NN_for_calibration}
\begin{figure}[h!]
	\includegraphics[trim={0.05cm 0.05cm 0.05cm 0.05cm},clip,width=0.75\textwidth]{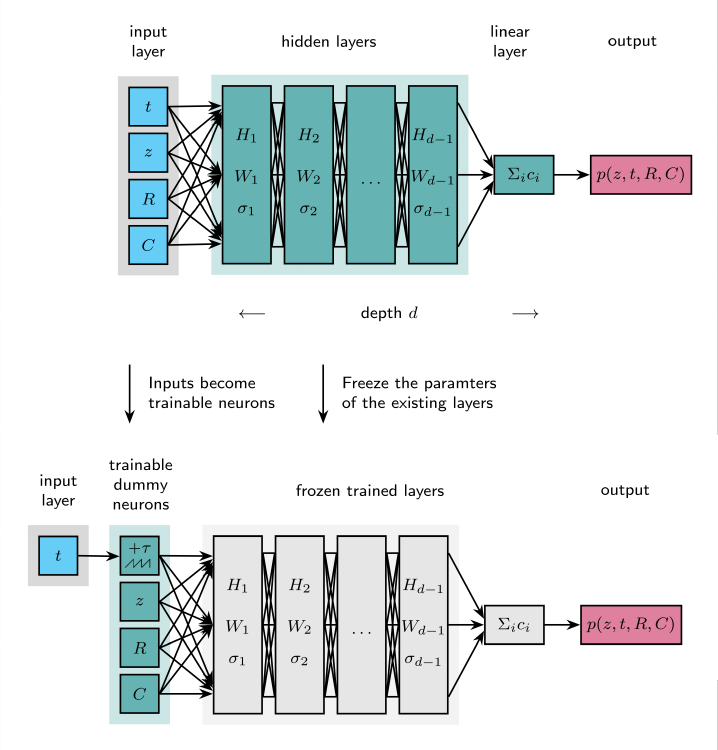}
	\caption{Scheme of the neural network surrogate and its extension to the calibration. Top part: general scheme of the fully-connected feed-forward neural network with input variables $t$, $z$, $R$ and $C$, multiple hidden layers $H_i$ (green) with parameters $W_i$ and activation functions $\phi_i$, and a linear layer, producing the output. This NN is trained to the reference data, to optimally reproduce the blood pressure in vessel 17. Bottom part: for the calibration, the pretrained NN is extended by four special neurons that incorporate the calibration parameters in its structure. The only remaining input is $t$ which is shifted and periodically backprojected by the subsequent new neuron. In the calibration, the parameters in the original layers (now grey) are frozen and solely the new parts are retrained to a specific measurement, yielding the calibration results.}
	\label{fig:NN_enhanced}
\end{figure}
We now propose an approach how to perform the calibration process directly in the framework of neural networks without employing an external optimization algorithm. As derived in Section \ref{sec:calibration}, the calibration can be expressed as the minimization of the cost functional $\hat{J}$ with respect to the variables $R$, $C$, $z$ and $\tau$. Our key idea is to embed the variables into the pretrained neural network by inserting special neurons (see Figure \ref{fig:NN_enhanced}). The calibration then is achieved by retraining only these special neurons to optimally match a given target measurement. Effectively, the trained NN will learn its own input to best reproduce a known output. This can avoid interface issues between the surrogate model and an external optimization tool.
Figure \ref{fig:NN_enhanced} illustrates how the calibration variables are incorporated into the NN: For each of the variables $z$, $R$ and $C$, we add a constant neuron at the front of the network and connect these solely to the respective input channels of the original model. These new constant neurons do not process any input but simply output fixed values stored in their "weights". This emulates a specific input to the original surrogate. By this, the modified NN depends only on the time variable $t$, while the others are transferred to its inner structure.

To incorporate the phase shift parameter $\tau$, we insert another special neuron, between the time input and the first hidden layer. This neuron solely adds a trainable bias to $t$, representing the time shift $\tau$. Optimizing this bias will provide the synchronization of the NN and the target measurement. Since this can push $t$ outside the interval $[0,T]$, we wrap the time using a sawtooth activation function with period $T$, to keep it within $[0,T]$. In total, the original neural network $f_{\vtheta}(z,t,R,C)$ is transformed into a modified version $f_{\mathbf{q}}(t)$, which is a function of only one variable, but with the four parameters $\mathbf{q}=(R,C,z,\tau)$:
$$
f_{\mathbf{q}}: [0,T] \rightarrow \R, \quad f_{\mathbf{q}}(t) = f_{\vtheta} \left( z, s_T(t+\tau), R, C_{tot} \,\right).
$$
Here, the sawtooth function is defined as $s_T(t)= T ( t/T - \lfloor t/T \rfloor )$. Evaluating $f_{\mathbf{q}}$ at a specific time and with fixed parameters $\mathbf{q}$ produces exactly the same output as evaluating the original surrogate at the appropriately shifted time $t+\tau$ (if necessary, projected onto $[0,T]$) and the corresponding inputs $z$, $R$ and $C$. For the actual calibration, the extended NN $f_{\mathbf{q}}(t)$ is provided with a series of $M$ reference pressure values $p_{17}(t_i)$, and is retrained exclusively for the parameters $\mathbf{q}$ until optimally matching the measurement. Similar to the initial training, the agreement is quantified by the mean square error
$$
\mathcal{L}(\mathbf{q}) = \frac{1}{M} \sum_{i=1}^{M} \left( f_\mathbf{q}(t_i) - p_{17}(t_i,z_i,R_i,C_i) \right)^2,
$$
which is minimized with respect to $\mathbf{q}$. Importantly, all previously trained parameters $\vartheta$ of the original model remain "frozen". Thus, we are only tuning the input of the general model, which is already informed about the dependencies on $R,C,z$ over some range, until it reproduces the desired measurement as best as possible. Once the training is complete, the inferred parameters $R_{opt}$ and $C_{opt}$ (as well as the inferred position $z$ and the phase-shift $\tau$) simply can be read from the trained special neurons. We stress that this embedded optimization is completely equivalent to minimizing the cost functional in the form of \ref{eq:costfun4p}, which is also a mean squared error. Moreover, it can be interpreted as an \emph{inverse} problem \cite{ying2022solving}: starting from a known output, the method traces back the optimal input. A key advantage of this approach lies in the fact that it uses the same optimization routine as in the primary training to the total reference data. This avoids additional implementation overhead and may improve computational efficiency, particularly when working with long measurement series (fine sampled and/or multi-period). Reusing the built-in optimizers, our NN-based calibration can exploit the weight-update routines of the NN framework, making it a fast and reliable technical implementation of the calibration problem.
\section{Numerical Tests of the Calibration}
\label{sec:tests}

We now analyze the NN-based calibration of the WK parameters when applied to several test cases. To this end, we use synthetic data -- specifically, data selected from the reference data set of Section \ref{subsec:data} -- as a proxy for real measurements. Although this may appear idealized, it offers clear advantages: It isolates the calibration method from real-world effects specific to the experimental setup and measurement device, allowing a focused evaluation of its core performance. We try to estimate these effects in a general way by Gaussian noise in one of the numerical studies below. A further benefit of using data from the existing reference set is that the true WK parameters and positions are known, enabling direct comparison with the calibrated results. In contrast, a potential disadvantage is that the NN may have seen the target measurement already during the primary training procedure (depending on the data thinning, see Section \ref{subsec:NN_less_data}). Effectively, this turns the calibration into a pure inverse recognition task, i.e. identify the associated input to a given, previously shown sample from a total set of training samples. To mitigate this, we run the calibration either with the reduced data versions or by presenting pressure curves with added noise to the NN. This will test less idealizing conditions, where the target was not exactly covered in the primary training, and hence provide insight into the generalizability to more realistic measurements.
\begin{figure}[h!]
\begin{minipage}{0.1\textwidth}
\end{minipage}%
\begin{minipage}{0.3\textwidth}
\centering
\includegraphics[clip, trim=0cm 0cm 0cm 0cm, scale=0.72]{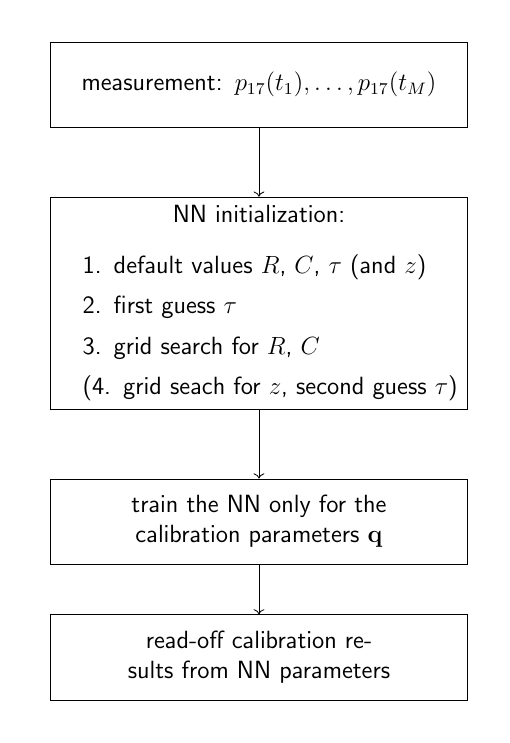}
\end{minipage}%
\begin{minipage}{0.5\textwidth}
\centering
\includegraphics[clip, trim=0cm 0cm 0cm 0cm, scale=0.35]{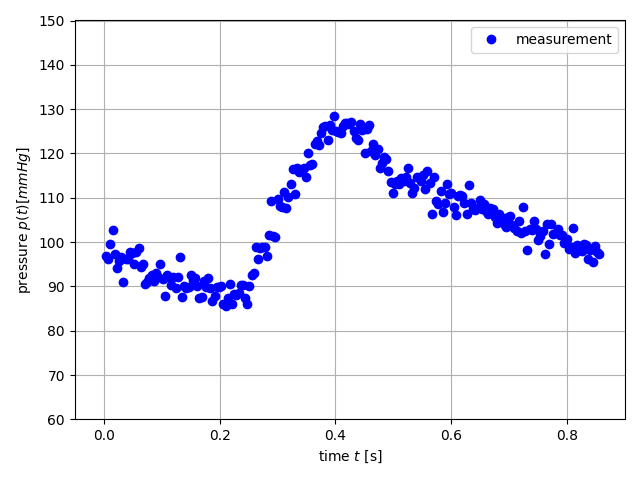}

\vspace{0.5em}
$\quad\downarrow$
\vspace{0.5em}

\includegraphics[clip, trim=0cm 0cm 0cm 0cm, scale=0.35]{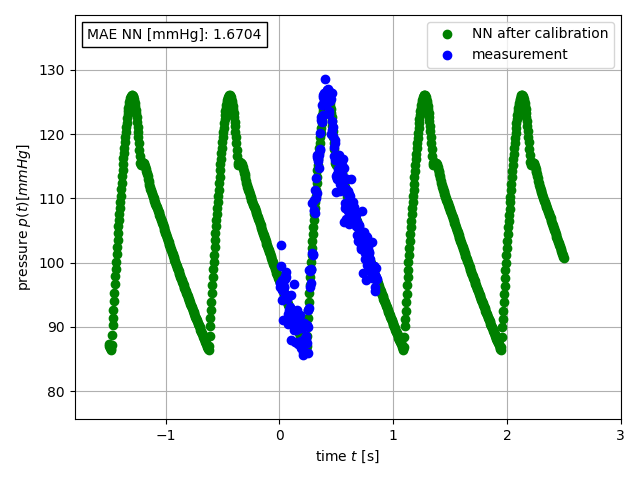}
\end{minipage}%
\begin{minipage}{0.1\textwidth}
\end{minipage}
\caption{Workflow of the calibration within the neural network. The neural network is retrained on a measurement (top right) by a multi-step procedure (left), to infer its optimal input parameters, corresponding to the optimal prediction (bottom right).}
\label{fig:workflow_calibration}
\end{figure}

\subsection{Calibration workflow}
The workflow of the calibration consists of several steps, as illustrated in Figure \ref{fig:workflow_calibration} (on the left), to adapt the pretrained and extended neural network to a given reference measurement (the same figure, on the right). The procedure begins with the initialization of the special neurons that represent the calibration variables
$\mathbf{q}$ (see Section \ref{subsec:NN_for_calibration}), which provides a strong starting point for the actual optimization. Due to the interdependence of the parameters, the initialization of the four dummy neurons is technically interwoven:
\begin{itemize}
\item The two Windkessel parameters $R$ and $C$ are initialized to their medians in their training range. The position $z$ is set to the measured value or, if there are no position data available, the mid-length of the vessel.
\item The time shift $\tau$ is then determined by synchronizing the diastolic pressure of the NN with the minimum of the measurement.
\item A quick $11\times11$ grid search provides an improved version of the WK parameters.
\item If the position is unknown, an 11-point grid search yields an updated guess for $z$. During this search, $\tau$ is also re-adjusted to maintain synchronization (due to the pulse propagation, reproduced by the NN, varying $z$ could break the alignment, as illustrated by Figure \ref{fig:phase_shift} in Section \ref{sec:calibration}).
\end{itemize}
Next, the Adam optimizer is applied to retrain neurons for the calibration variables $\mathbf{q}$. Here, we follow a two-stage learning rate schedule ($\eta_1=0.01$ for 500 epochs, then $\eta_2=0.001$ for 1500 epochs), and optimize all calibration variables jointly. The optimizer does not split the data into batches, but processes the $M$ time points at once, which is feasible, since $M$ is of moderate size (in our case 228), and reflects that all $M$ points form a unit. The final calibration results are simply retrieved from the respective neurons, yielding the most plausible combination of the WK parameters, phase shift, and, if necessary, position. A graphical analysis serves as a final cross-check of the match between predicted and measured blood pressure.
For the following, we denote by $\tilde{R}$ and $\tilde{C}$ the ground truth WK parameters of the underlying simulation and by $R_{cal}$ and $C_{cal}$ the results of the calibration (generally, the tilde symbol marks the true value). For brevity, we usually skip the units $\uR$ for resistance and $\uC$ for the compliance. In addition to that, we evaluate the calibration results in terms of relative/percentage errors, defined as
$$
\frac{R_{cal}-\tilde{R}}{\tilde{R}}, \quad \frac{C_{cal}-\tilde{C}}{\tilde{C}} \quad\quad \text{(in \%)}.
$$
The other two optimized parameters, denoted by $z_{opt}$ and $\tau_{opt}$, are evaluated only in absolute numbers -- relative errors for values close to 0, which can occur for $\tau$ and $z$ are not comparable. The ground truth position, denoted by $\tilde{z}$, is given by the position of the simulation; for $\tau$, we avoid stating a true value in lack of a precise definition of the minimum beyond the sampling rate. Most of the parts consider recovering the WK parameters for two test cases, which can be thought of as hypothetical subjects: Case $1$ has the parameters $\tilde{R}_1=1.34$ and $\tilde{C}_1=0.945$, which are the medians of the simulated parameter range $Q$, Case $2$ is defined by the pair $\tilde{R}_2=1.441$ and $\tilde{C}_2=0.890$, located in the outer region of $Q$. However, we consider different true positions $\tilde{z}_1$ and $\tilde{z}_2$ in the parts below, depending on the specific analysis.

\subsection{Calibration on a Pulse Wave Signal at Fixed Position}
Our baseline test assesses the most idealized scenario: The measurements are taken from the reference data set with their original phase shift, and their position $\tilde{z}$ is known. Furthermore, we apply the full-data variant of the NN, without a reduced resolution of the reference data, carrying the full information on the training set in its pretrained weights. Thus, the pretrained NN effectively has to simply recover the underlying WK parameters of one of its training curves, which we described above as inverse recognition. The results, presented in Table \ref{tab:calibration_test_1}, demonstrate that the approach accurately identifies the true WK parameters of the two target cases, regardless if the true position $\tilde{z}$ is close to the inlet, in the middle of the artery or close to the outlet. In detail, we observe that the determined initial values are highly reasonable (in the first case, the initialization is exceptionally close to the target, since the target is very close to one of the grid points). After optimization, all relative errors are less than $\pm 0.07\%$, providing a solid proof of principle of our method.

The accuracy is especially satisfying if one takes into account that the mesh widths of $R$ and $C$ in the simulated data are ca. $0.9 \%$ on relative scales. One observes a slightly better precision for the calibrated resistance, which follows from its larger influence on the pressure curves. The compliance has smaller effects, making it more difficult to trace back changes in the pressure curves to this parameter. Obtaining differences in precision is then expectable at a joint optimization, and, in fact, this will be observable throughout the further tests. Computationally, the calibration procedure is finished after seconds on a standard notebook due to the NN's compiled graph. This baseline test successfully validates our methodology for solving the minimization problem with a properly trained NN surrogate.

\begin{table}[H]
\centering{}%
\renewcommand{\arraystretch}{1.3}
\begin{tabular}{|c||c|c||c|c|c|c|}
\hline 
\multicolumn{7}{|c|}{Case 1: $\tilde{R}=1.34$, $\tilde{C}=0.945$}\tabularnewline
\hline 
$\tilde{z}$  & $R_{in}$  & $C_{in}$  & $R_{cal}$  & $C_{cal}$  & $\Delta R$  & $\Delta C$ \tabularnewline
\hline 
$4.22 \, \unit{cm}$  & 1.3401 & 0.9449 & 1.3401 & 0.9449 & 0.004 \% & -0.006 \%\tabularnewline
\hline 
$21.10 \, \unit{cm}$  & 1.3401 & 0.9449 & 1.3401 & 0.9451 & 0.007 \% & 0.015 \%\tabularnewline
\hline 
$37.98 \, \unit{cm}$  & 1.3401 & 0.9449 & 1.3401 & 0.9448 & 0.011 \% & -0.021 \%\tabularnewline
\hline 
\end{tabular}\vspace{0.5em}
\renewcommand{\arraystretch}{1.3}
\begin{tabular}{|c||c|c||c|c|c|c|}
\hline 
\multicolumn{7}{|c|}{Case 2: $\tilde{R}=1.441$, $\tilde{C}=0.890$}\tabularnewline
\hline 
$\tilde{z}$  & $R_{in}$  & $C_{in}$  & $R_{cal}$  & $C_{cal}$  & $\Delta R$  & $\Delta C$\tabularnewline
\hline 
$4.22 \, \unit{cm}$  & 1.4367 & 0.8994 & 1.4412 & 0.8903 & 0.011 \% & 0.035 \%\tabularnewline
\hline 
$21.10 \, \unit{cm}$  & 1.4367 & 0.8540 & 1.4411 & 0.8905 & 0.009 \% & 0.061 \%\tabularnewline
\hline 
$37.98 \, \unit{cm}$  & 1.4367 & 0.8994 & 1.4412 & 0.8896 & 0.015 \% & -0.041 \%\tabularnewline
\hline 
\end{tabular}
\caption{Baseline test for the calibration: Recovery of the WK parameters from one of the reference curves of the training data, at various, known positions, and with equal phase as captured by the NN. Listed are the first guesses obtained from the initialization procedure and the actual calibration results, using the full-data NN.}
\label{tab:calibration_test_1} 
\end{table}

\subsection{Phase synchronization}
\label{subsec:phase_synchronization}
\begin{figure}[h!]
\begin{minipage}{0.2\textwidth}
\framebox[3.5cm]{\textsf{Default parameters}}
\,\\
\end{minipage}%
\begin{minipage}{0.8\textwidth}
\centering
\includegraphics[clip, trim=0cm 0cm 0cm 0cm, scale=0.5]{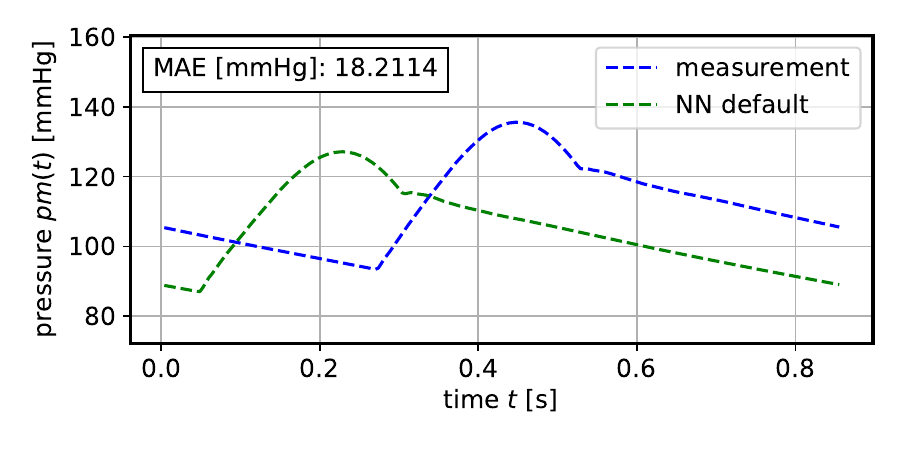}
\end{minipage}

\begin{minipage}{0.2\textwidth}
\framebox[3.5cm]{\textsf{Initial guess for $\tau$}}
\,\\
\end{minipage}%
\begin{minipage}{0.8\textwidth}
\centering
\includegraphics[clip, trim=0cm 0cm 0cm 0cm, scale=0.5]{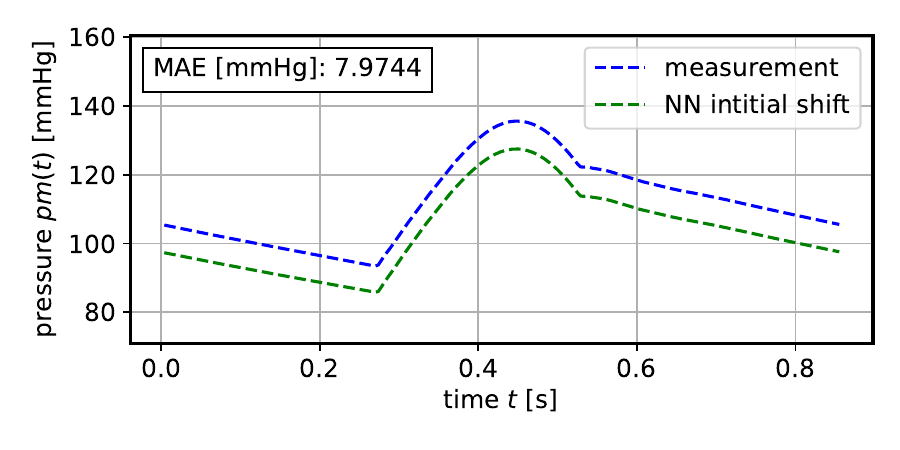}
\end{minipage}

\begin{minipage}{0.2\textwidth}
\framebox[3.5cm]{\textsf{After calibration}}
\,\\
\end{minipage}%
\begin{minipage}{0.8\textwidth}
\centering
\includegraphics[clip, trim=0cm 0cm 0cm 0cm, scale=0.5]{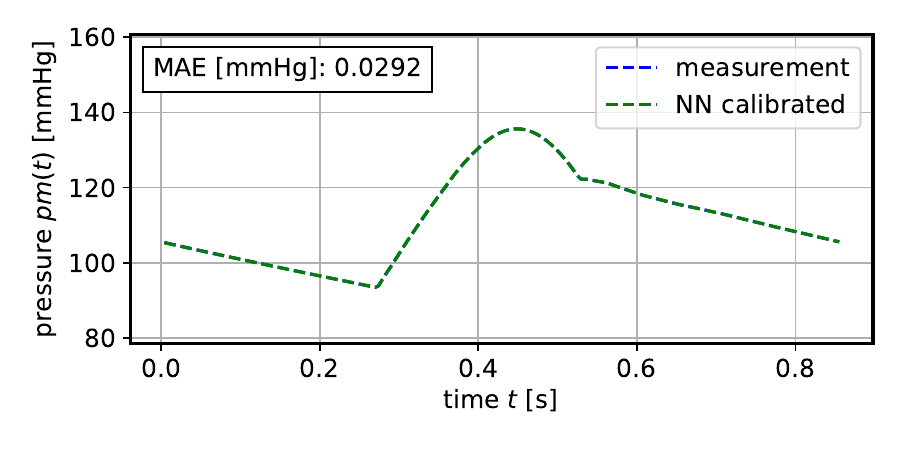}
\end{minipage}

\caption{Stages of the phase synchronization with the neural network during the calibration procedure. All three plots show a reference curve, with the systolic pressure shifted to ca. $0.6\;\unit{s}$ and the NN prediction at the different steps (top: default state, middle: after first guess for $\tau$, bottom: after calibration.}
\label{fig:calibrate_with_phase_shift_2}
\end{figure}

Different from above, the measurement time series of the pressures $p_{17}$ can start with a different phase, e.g. due to a different convention or even the absence of a clear convention. Figure \ref{fig:calibrate_with_phase_shift_2} shows by means of an example (Case 2 at $\tilde{z}=37.98$ cm), how the extended NN adapts to a time-shifted reference curve. Here, the measurement series was "rolled" in time, which here shifts, for instance, the systolic pressure to $t=0.45 \, \unit{s}$. The default state of the NN, with $\tau=0 \, \unit{s}$, is not synchronized. The first guess for $\tau$ within the initialization routine then provides a basically synchronous, but uncalibrated prediction. The subsequent optimization jointly recovers the optimal WK parameters and phase shift. We remark that treating the shift parameter $\tau$ as a continuous variable is advanced over rolling the order of the predicted pressures until synchronization, since the latter approach can cause discretization artifacts.

To confirm that this works in general, we repeated two calibration scenarios from above at known positions (Case 1 close to the inlet, Case 2 close to the outlet) at varying measurement shifts. We systematically rolled the measurement series by 10 time points between each run, effectively shifting the phase of the systolic pressure $t_{max}$ (and all other features of the pulse wave) across the total interval $[0,T]$.

\begin{figure}[h!]
\begin{minipage}{0.3\textwidth}
\begin{center}
\includegraphics[clip, trim=0cm 0cm 0cm 0cm, scale=0.5]{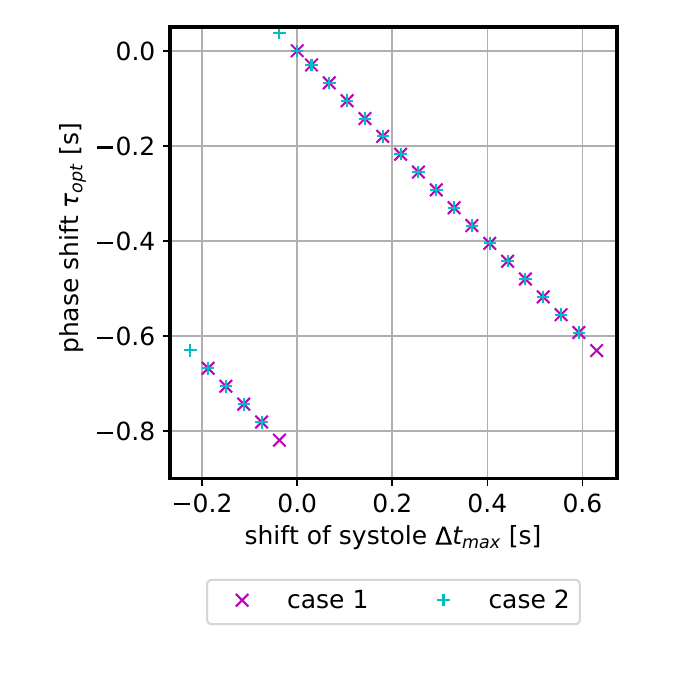}
\end{center}
\end{minipage}%
\begin{minipage}{0.7\textwidth}
\begin{center}
\includegraphics[clip, trim=0cm 0cm 0cm 0cm, scale=0.5]{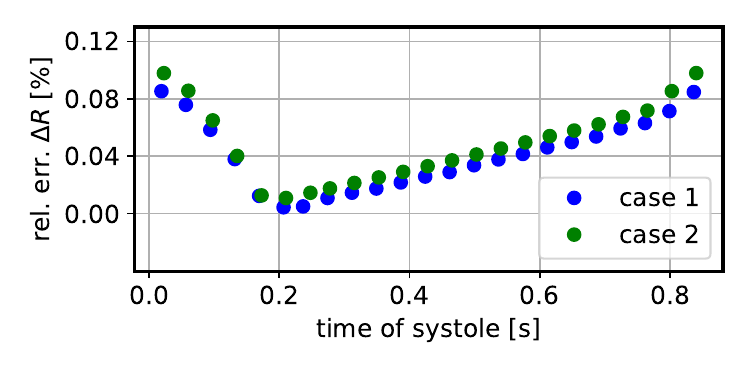}\\
\includegraphics[clip, trim=0cm 0cm 0cm 0cm, scale=0.5]{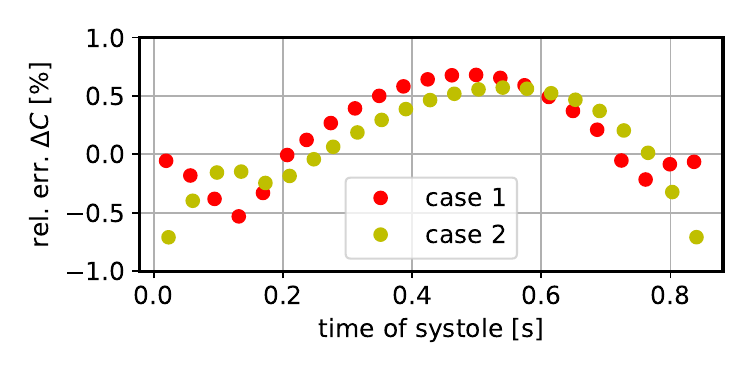}
\end{center}
\end{minipage}
\caption{Calibration at varying phase of the measurement. Left: Optimized phase shift vs. time shift of the systolic pressure in the reference curve, Right: Percentage error of times the time point of the systolic pressure in the reference curve vs. $R$ (top) and $C$ (bottom). Case $1$ refers to $\tilde{z}=4.22 \, \unit{cm}$ and Case $2$ refers to $\tilde{z}=37.98 \, \unit{cm}$.}
\label{fig:calibrate_with_phase_shift}
\end{figure}

The results, presented in Figure \ref{fig:calibrate_with_phase_shift}, shows the optimal phase shift $\tau_{opt}$ versus the 
time shift of the systole with respect to the original curve (top), and the percentage errors of the calibrated $R$ and $C$ values against the time point of the systole in the reference curve (original time points at ca. $0.21 \, \unit{s}$ in Case 1 and ca. $0.25 \, \unit{s}$ in Case 2). Our method successfully identifies the correct phase shift in all cases, compensating for the shift in the measurement. However, the accuracy of the obtained calibrated WK parameter varies. The lowest errors occur at smaller shifts; for larger shifts, the precision decreases. This is likely a boundary effect, caused by ambiguities in the cut-off of the periodic window (since the accuracy of period is limited by the sampling rate). Surprisingly, the percentage error of $R$ seems to have inherited the negative profile of the pressure curve. We assume that if the largest pressure values, i.e. close to the systole, fall into the cut-off region, small variations at the boundary produce relatively large changes in mean square error, and that these are predominantly assigned to $R$.

In any case, the resistance $R$ is recovered in a precise way with percentage errors of less than $0.12$\%. In contrast, the compliance $C$ has remarkably larger percentage errors, falling within the range of ca. $\pm 1$ \%. We assign this difference to the weaker influence of $C$ on the pressure curve, and even more to the time-dependence of these effects, which means that even small temporal offsets between the NN and the reference can lead to a relatively large error.

Despite this shortcoming, our method successfully determines time synchronization and provides highly reliable calibration of resistance. While calibration of the compliance is somewhat less precise, the results are still plausible. Our approach works best when the starting phases of the measurement and the NN are close to each other, but it can still effectively correct for significant mismatches. For future work, it would be interesting to consider multi-period data which might contain incomplete periods but be less affected by boundary effects. Based on these findings, the remainder of the tests will use the convention that the measured curve starts at the minimum (the diastolic pressure), as this is realistic for actual measurements and requires only for a small phase correction.

\subsection{Uncertainty Analysis for the Position}

In clinical or experimental settings, the measurement position $\tilde{z}$ can be difficult to determine in a precise way. Since our neural network surrogate relies on this information, any measurement uncertainty in $\tilde{z}$ will impact the calibration results. To quantify this influence, we analyzed how the calibrated parameters $R$ and $C$ change when the NN is evaluated at an incorrect position, deviating from the true position in a certain window.

For this purpose, the presumed position was varied within a range of $\pm 2 \, \unit{cm}$ around the exact position, which is $\tilde{z}=21.1 \, \unit{cm}$ in this setup. The size of the uncertainty window is reasonable for actual measurements. Then the calibration was run at each $0.2 \, \unit{cm}$ step in that range, where the position in the NN is kept constant throughout each run, i.e. not included in the optimization. By this, we can estimate the isolated error caused by the more or less inaccurate assumption on the measurement. Importantly, the reference curve was the same, i.e. at the true position, across the runs, with the convention to start at the diastole.

\begin{figure}[h!]
	\centering
\begin{minipage}{0.5\textwidth}
\centering
\textsf{case 1}\\
\includegraphics[width=0.85\textwidth]{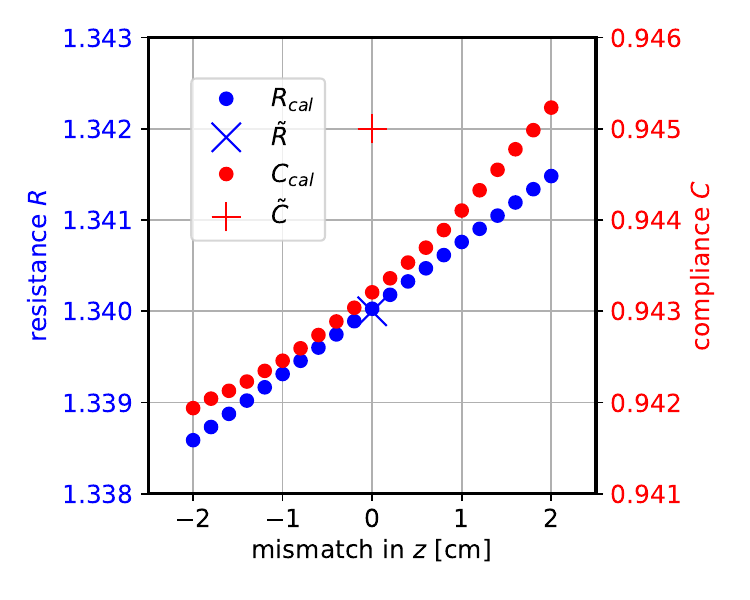}
\end{minipage}
\begin{minipage}{0.5\textwidth}
\centering
\textsf{case 2}\\
\includegraphics[width=0.85\textwidth]{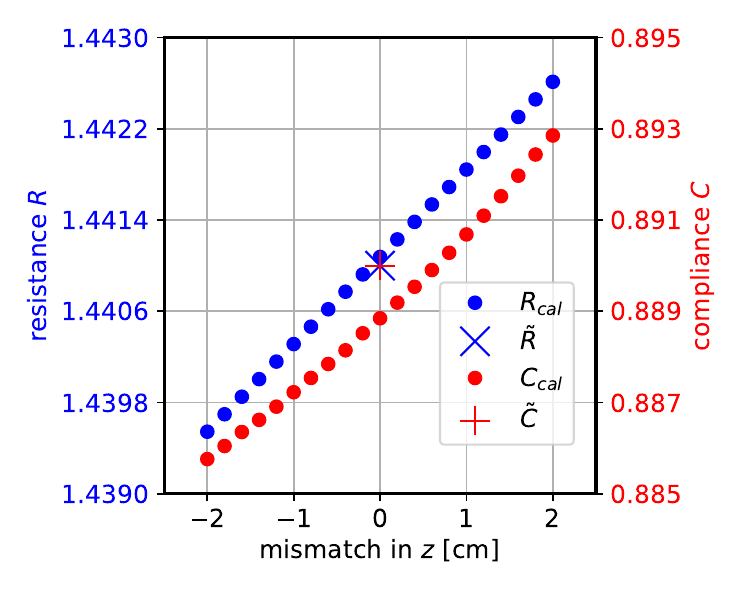}
\end{minipage}
\caption{\label{fig:uncertainty_z} Influence of the uncertainty in the measurement position on the calibration result for the two test cases. The NN assumes a fixed position which deviates by up to $\pm 2 \, \unit{cm}$ from the true position $\tilde{z}=l/2$, yielding the plotted calibrated values. The crosses indicate the underlying true WK parameters.}
\end{figure}

Figure \ref{fig:uncertainty_z} shows the position mismatch (left: Case $1$, right: Case $2$) vs. the calibrated resistances and compliances. Both the resistance and the compliance are basically monotonically influenced by the mismatch in assumed and actual position within a small range, demonstrating the robustness of our approach to imprecise position measurements. The calibrated values for $R$ deviate by less than $\pm 0.002$ (in $\uR$) from the true value within the tested position range, and these deviations are centered around the true value. While the calibrated values for $C$ show a slight negative bias -- they are consistently underestimated at the true position -- this appears to be related to the new convention of the signal start (at diastole) and the higher sensitivity of $C$ to small changes in the phase to the phase synchronization, see Section \ref{subsec:phase_synchronization}. It cna be observed that the optimization of $\tau$ is important in this setup, since it compensates the phase shift due to the pulse propagation which comes into play when the NN is evaluated at a different position. Again regarding to $C$, the uncertainty in $\tilde{z}$ itself introduces a very small error of at most $\pm 0.005$ (in $\uC$) over the tested position range.

In conclusion, our calibration methodology tolerates a moderate uncertainty in the position with only minor and well-controlled additional error in the calibrated WK parameters. This is helpful for practical applications, where e.g. ambiguities in a reference point due to anatomical differences and the dimensions of the measurement device impair determining the position exactly.

\subsection{Calibration at Unknown Position}

Instead of considering an imprecise measurement of position, we now investigate how our approach performs when the position $\tilde{z}$ is entirely unknown. As explained in Section \ref{sec:calibration} the underlying minimization problem is easily extended by the variable $z$. In this case, the pretrained neural network, which captures the spatial dependencies along the artery, is used to simultaneously infer the optimal position along with the WK parameters and the phase shift.

To evaluate this capability, we ran the calibration for both test cases across all 31 available positions in the interval $[0,L]$. The extended NN starts from the default position $z=l/2$ and optimizes all parameters jointly, following the workflow presented above. Importantly, all 31 reference curves are aligned to begin at their diastolic pressure, thereby eliminating information on the phase, which could easily reveal $z$. Thus, the NN must rely on the remaining, more subtle effects of $z$ -- such as changes in amplitude and shape of the pulse wave -- to infer the position.

It can be seen from Table \ref{tab:calibration_unknown_position} that the optimal positions are inferred accurately; they only slightly underestimate the actual positions by at most $0.2 \, \unit{cm}$. The same table confirms that the Windkessel parameters are estimated accurately as well: resistance values are as precise as in the known-position case (within $\pm 0.002 \cdot \uR$), while the compliance shows slightly higher deviation (up to $0.004 \cdot \uC$), but remains on a satisfactory level. This error, which has a negative bias, may be attributed to the phase convention and its correction (as discussed earlier), and also to the similarity between the effects of $z$ and $C$ on the pressure waveform -- making it more difficult for the NN to disentangle their influence during optimization.

In summary, this analysis demonstrates that our method can perform calibration highly accurately even without prior knowledge of the measurement position. The quality of the estimated Windkessel parameters remains comparable to that achieved by optimizing at imprecise position measurement. Thus, rather than speculating on a position, the method is capable of directly inferring $z$. However, we note that the recovery of $z$ relies on subtle variations in the pressure curve, and thus requires high-precision measurements to work well in practice.

\begin{table}[h!]
\centering{}%
\renewcommand{\arraystretch}{1.2}
\begin{tabular}{|c|c||c|c|c|c|c|}
\hline 
 &  & $\Delta z$ [cm] & $R_{cal}$ & $C_{cal}$ & $\Delta R$ & $\Delta C$\tabularnewline
\hline 
\hline 
\multirow{2}{*}{Case $1$} & min  & -0.1647 & 1.3399 & 0.9420 & -0.009 \% & -0.315 \%\tabularnewline
\cline{2-7}
 & max  & -0.0076 & 1.3401 & 0.9439 & 0.008 \% & -0.113 \%\tabularnewline
\hline 
\hline 
\multirow{2}{*}{Case $2$} & min  & -0.1912 & 1.4409 & 0.8869 & -0.005 \% & -0.351 \%\tabularnewline
\cline{2-7}
 & max  & -0.0147 & 1.4411 & 0.8894 & 0.007 \% & -0.070 \%\tabularnewline
\hline 
\end{tabular}
\caption{Calibration at unknown position. The table reports the minimal and maximal deviations of the calibration when varying the true measurement position over $[0,L]$ at all sampled positions.}
\label{tab:calibration_unknown_position} 
\end{table}

\subsection{Sensitivity to Noise}

In view of real-world measurements, it is crucial to understand the sensitivity of our method to noise in the reference pressure curve and how distortions in the data affect the stability of the calibration results. To estimate the effects, we repreated the calibration for the two test cases by adding uncorrelated Gaussian noise to the synthetic reference curve, i.e. to perform the calibration on
$$
p_{17} (z,t,R,C) \; \mapsto \; p_{17} (z,t,R,C) + \mathcal{N}(\mu,\sigma),
$$
at various values for the bias $\mu$ and the standard deviation $\sigma$. The characteristics of uncorrelated Gaussian noise are a reasonable model for the errors caused by the measurement device. Physiological fluctuations typically exhibit temporal correlations or slower changes (e.g., variations in a pulse wave, variations of the heart frequency), and require a more sophisticated noise model. Hence, our additive noise tends to overestimate the variability in real measurements.

In our numerical study, we consider biases within $\mu \in \left\{ 0,\,1,\,2  \right\}$ $\left(\mu \text{ in } \unit{mmHg}\right)$ and standard deviations $\sigma \in \left[0,5\right]$ $\left(\sigma \text{ in } \unit{mmHg}\right)$ consistent with established precision standards for clinical continuous blood pressure sensors \cite{stergiou2018universal,wang2022novel,ISO81060}. Since the mean arterial pressure in the considered vessel is about $100 \, \unit{mmHg}$, the above values for $\mu$ and $\sigma$ can also be interpreted as percentage error strengths. For each noise configuration, we performed a Monte Carlo study over 50 independent calibration runs with newly generated noise in each iteration. As before, the target parameters are the test cases $1$ and $2$, where the measurement position is at $\tilde{z}=l/2$, which is considered as known and not optimized, and the measurements are aligned to start at the beginning of the diastole. We refer back to Figure \ref{fig:workflow_calibration} on the right, illustrating that the NN can adapt to a noisy measurement.

\begin{figure}[h!]
	\centering
\begin{minipage}{0.5\textwidth}
\includegraphics[width=0.90\textwidth]{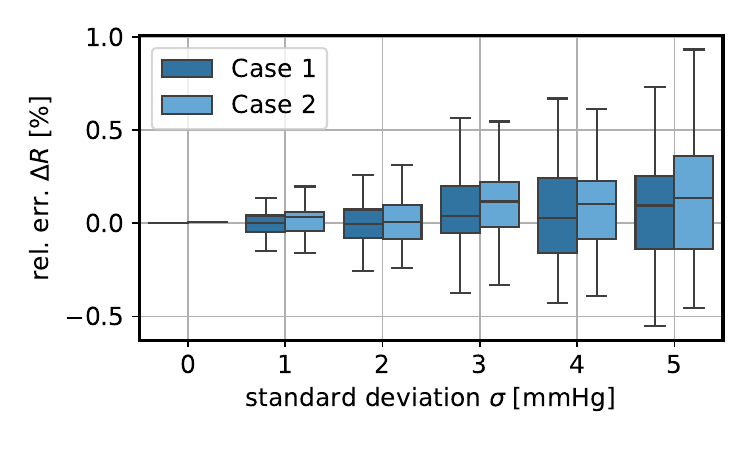}
\end{minipage}
\begin{minipage}{0.5\textwidth}
\includegraphics[width=0.90\textwidth]{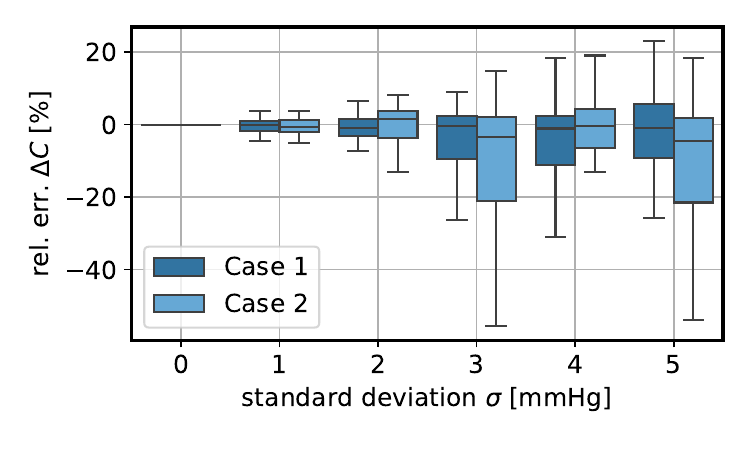}
\end{minipage}
\begin{minipage}{0.5\textwidth}
\includegraphics[width=0.90\textwidth]{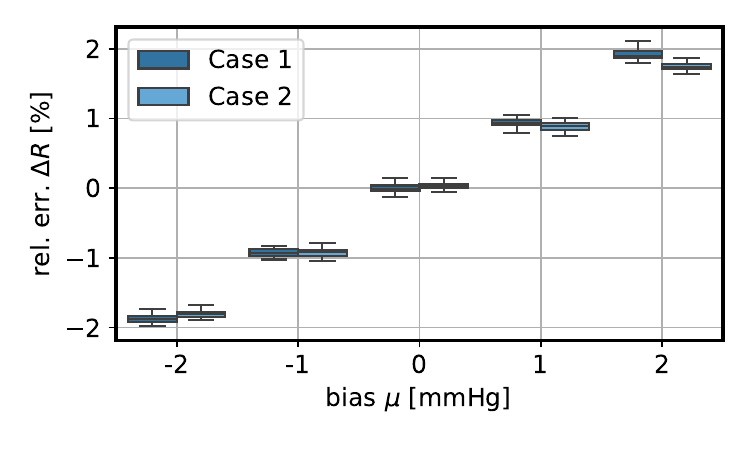}
\end{minipage}
\begin{minipage}{0.5\textwidth}
\includegraphics[width=0.90\textwidth]{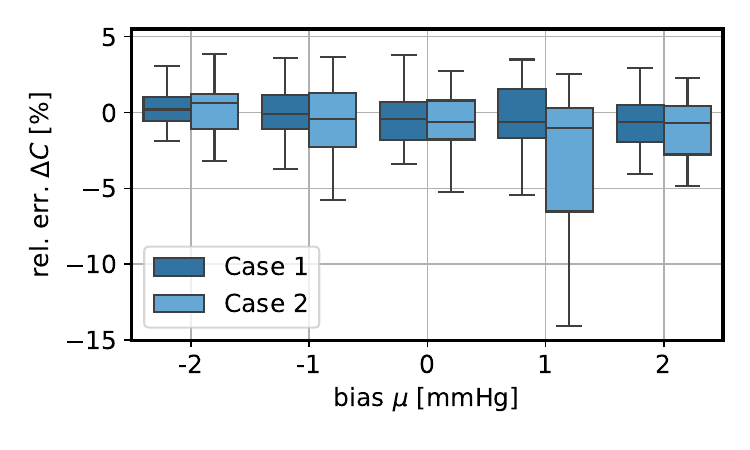}
\end{minipage}
\caption{\label{fig:noisy_1} Error distributions of the calibrated WK parameters with noisy measurements. The box plots visualize the distribution of the percentage error of the optimal resistance (left diagrams) and compliance (right diagrams) obtained by a Monte Carlo study with $50$ calibration runs for each of the two test cases at varying additive Gaussian noise $\mathcal{N}(\mu,\sigma)$. Upper panel: Variation of standard deviation $\sigma$ at zero bias, lower panel: Variation of bias $\mu$ at constant standard deviation of $1 \, \unit{mmHg}$.}
\end{figure}

The results of the actual numerical study are shown in the boxplots of Figure \ref{fig:noisy_1}, which represent the distributions of the percentage error of $R$ and $C$ after calibration. The upper plots correspond to bias-free noise with increasing variance, while the lower plots correspond to varying bias at a fixed low standard deviation of $1 \, \unit{mmHg}$. 

In the bias-free cases, increasing $\sigma$ leads to broader distributions of the optimal parameters, in particular for the compliance. The resistance remains stable within $\pm 1 \%$ even at $5 \, \unit{mmHg}$, while the compliance becomes unrecoverable for $\sigma \geq 2 \, \unit{mmHg}$. This aligns with the known properties of the dataset: pressure curves are more weakly and time-dependently influenced by $C$ (see Fig. \ref{fig:partial_dependencies_p}). Even varying $C$ across its full range only results in pressure differences of about $2 \, \unit{mmHg}$. Hence, accurate calibration of $C$ requires very high-precision measurements, whereas $R$, due to its larger global, time-independent effect on the pressure, is more robust to bias-free noise. Here, one has to consider that the noise will cancel out statistically over one period, and since the effect of $R$ is basically a global shift, proportional to $R$, the average effect will be very close to the noiseless case. In that sense, we expect $R$ to be well-recoverable even under use of time-unresolved measurements like classical SP/DP data. We also note that the calibration results under bias-free Gaussian noise can be interpreted as maximum likelihood estimators since minimization of the mean square error given by $\hat{J}$ under Gaussian noise assumption is equivalent to maximum likelihood estimation \cite{bishop2006pattern,wang2023intuitive}. Moreover, modeling the blood pressures by a Gaussian model with NN-predicted means and additional variance representing measurement errors could further quantify uncertainty \cite{wang2023intuitive}.

The situation differs when the noise includes a non-zero bias. In this case, the results for the resistance are sharply distributed but are consistently offset from the true value, in proportion to the bias $\mu$. This matches the expectation from the data: Since a change of resistance mainly shifts the pressure curve proportionally, a constant bias will be misattributed to an accordingly smaller or larger $R$ (which would lead to the biased curve). The effect of the bias can be estimated analytically, if one assumes a linear relation between $R$ and the pressure $p_{17}$. It can be taken from Figure \ref{fig:partial_dependencies_p} (bottom left), that the pressure changes by $22 \, \unit{mmHg}$ if $R$ varies on its full simulated range (from its minimum to its maximum), on a range of width $0.268 \cdot \uR$. Thus, the slope can be approximated by
$$
\frac{\partial p}{\partial R} \approx \frac{22\;\unit{mmHg}}{0.268 \cdot \uR} \approx 80 \, \frac{\unit{mmHg}}{\uR}.
$$
Inverting this, yields the error propagation for $\mu= \pm 2 \, \unit{mmHg}$ on $R$:
$$
\Delta R \approx \left( \frac{\partial p}{\partial R} \right)^{-1} \mu = \pm 0.0125 \cdot 2 \, \unit{mmHg} = \pm 0.025 \cdot \uR.
$$
This matches the numerical study since relative errors of $\pm 2\%$ translate to absolute errors of about $\pm 0.03 \cdot \uR$ for the two cases. For the compliance, bias appears less detrimental than variance -- possibly because the relevant time-dependent features of the pressure curve are not disturbed.

In summary, accurate calibration requires very precise data: recovering $R$ requires low bias, covering $C$ low variance. Noise amplitudes up to ca. $1 \, \unit{mmHg}$ appear tolerable, but offsets should be kept very close to zero. Generally, while noisy data introduces uncertainty, this seems to result from the reduced data quality itself and not from the optimization method. The minimization of $\hat{J}$ naturally averages out symmetric noise without the need for noise filtering or reduction. Still, the effects of $C$ might be too subtle to be resolved by a realistic, noisy measurement. If accurate calibration of $C$ is critical, alternative setups of the cardiovascular network with more pronounced effects of $C$ or transient dynamics should be investigated.
 
\subsection{Calibration with Reduced-Data Models}
\label{subsec:test_calibration_less_data}
This part analyzes how reducing the resolution in the NN's training data indirectly affects the calibration of the WK parameters. To this end, we repeated the calibration for the two test cases while using the various reduced-data models from Section \ref{subsec:NN_less_data}, in which the sampling meshes in time, position, resistance and compliance were isolated or jointly reduced. The calibration assumes fixed positions, i.e., $\tilde{z}=0 \, \unit{cm}$ for Case $1$  and $\tilde{z} = 23.96 \, \unit{cm}$ for Case $2$. Importantly, the target parameters of Case $1$ lie almost exactly on a grid point on $R$ and $C$ on the rougher meshes, while those of Case $2$ mostly fall outside of it. Thus, Case $1$ allows to estimate if the NN calibration is influenced by training data in the close surrounding of the target: if removing the direct neighborhood, but not the target itself, improves the calibration, this would suggest a higher specificity of the model, if it worsens the calibration, this would suggest transfer learning. Case $2$, in contrast, can provide insight into the interpolating capability of the calibration. The same distinction between the cases applies spatially: Case $1$ is generally concentrated on the grid, while Case $2$ considers targets that are always off-grid. To better explore models trained on a single position, we applied the NN trained solely at the center (as before) and additionally two versions trained solely at the inlet (z0) or the outlet (zl). The time resolution of the reference measurement remains fixed at $M=228$ steps across all tests. Consequently, the NN is also evaluated on all these steps to calculate the pairwise differences in 
$\hat{J}$, even if the respective variant was trained at a smaller time resolution. Thus, one cannot distinguish clearly between included and excluded (interpolating) cases for the time variable, as it applies for $z$, $R$, and $C$ by choice of the target cases, which are either fully on or off the mesh in these variables.

\begin{figure}[h!]
\begin{minipage}{0.7\textwidth}
\includegraphics[clip, trim=2cm 0cm 2cm 0cm, scale=0.47]{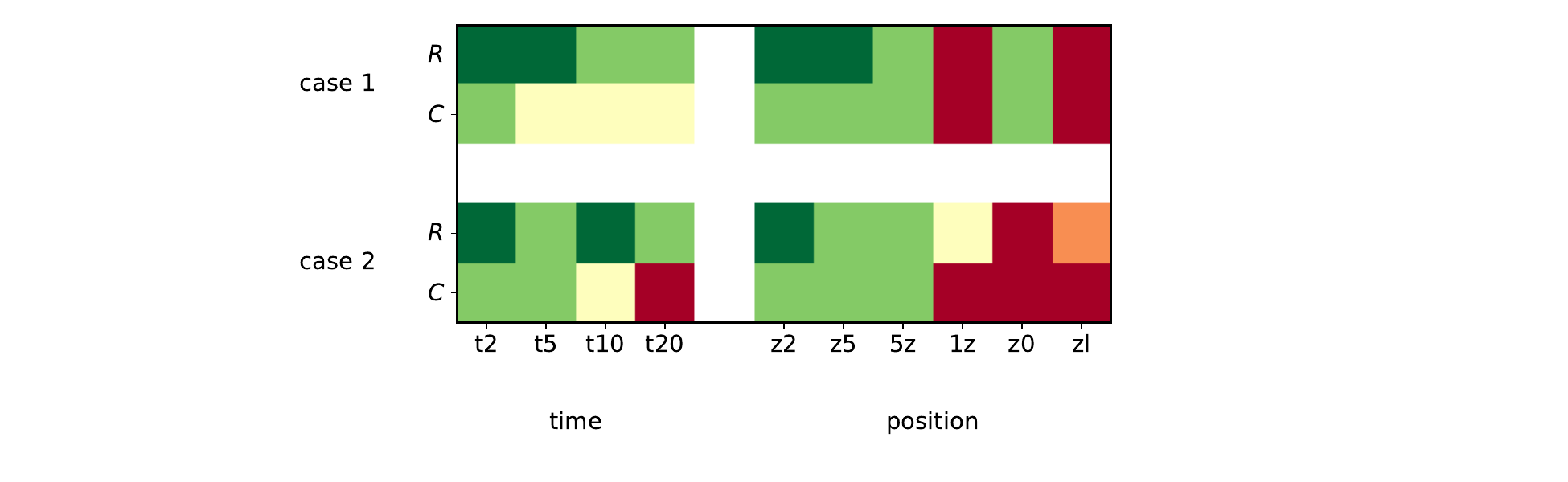}
\end{minipage}%
\begin{minipage}{0.3\textwidth}
\includegraphics[scale=0.5]{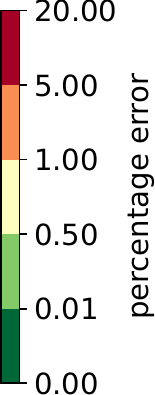}
\end{minipage}
\includegraphics[scale=0.40]{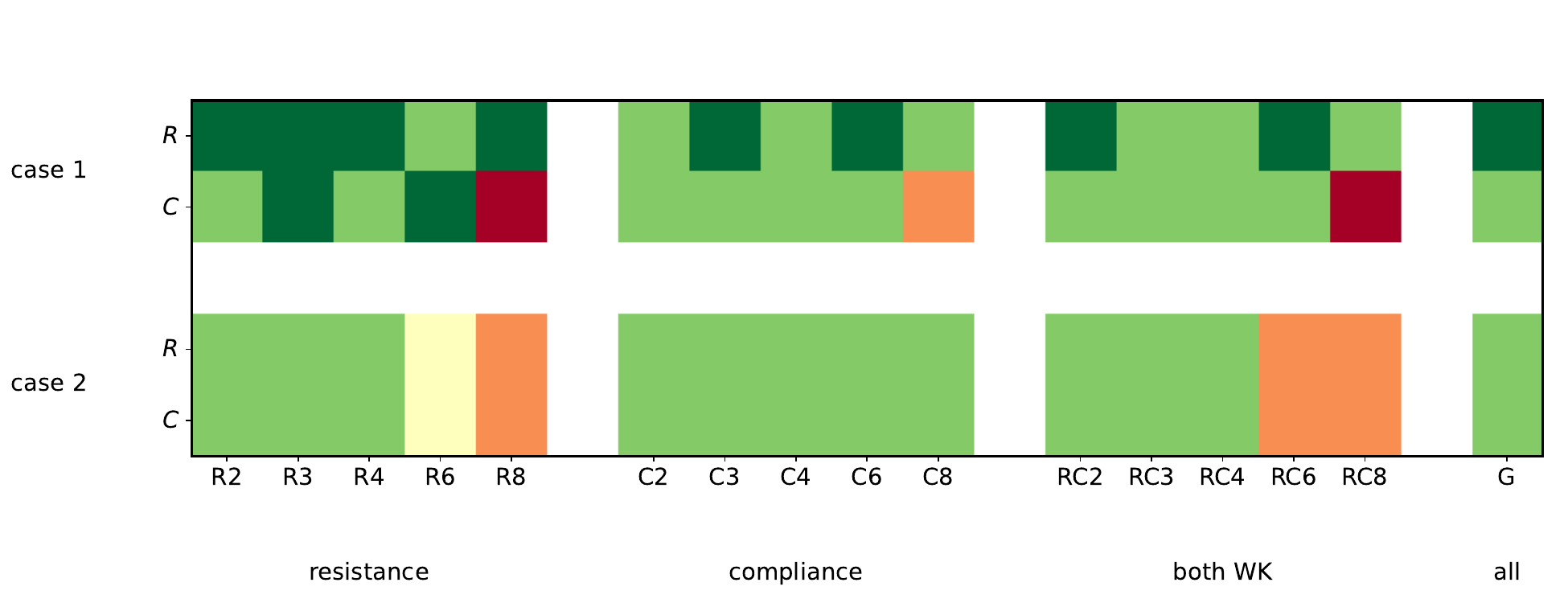}
\caption{\label{fig:reduced_data_calibration} Calibration accuracy of the WK parameters using reduced-data models. The heatmaps show the percentage error of the optimal resistance and compliance for Case $1$ at $\tilde{z}=0 \, \unit{cm}$ and Case $2$ at $\tilde{z}=23.96 \, \unit{cm}$ for the considered variants of training data thinning for the underlying NN.}
\end{figure}

Figure \ref{fig:reduced_data_calibration} shows the values of the percentage errors of the calibrated WK parameters for the two test cases with respect to different reduction variants. Color levels in the heatmaps mark the accuracy: dark/light green for high (percentage error $<0.5 \%$), yellow for acceptable ($\geq 0.5 \%$ but $< 1 \%$), and orange/red for poor ($\geq 1 \%$). Overall, the tolerable mesh coarsening aligns well with the findings at the model level (Section \ref{subsec:NN_less_data}). The calibration results are robust up to a factor of 5 to 10 in time and space, and up to a factor of 4 for $R$ and $C$, also if both parameters are coarsened jointly. Notably, the global thinning variant, which is trained on just $0.4 \%$ of the orginal data set, still achieves satisfactory errors of less than $\pm 0.7 \%$.

More in detail, training only on five of the 31 positions on the position grid (inlet, center, outlet, two positions between) performs well. Single-point models only provide reasonable values when the measurement and training positions match (this occurs for case 1 with model variant z0, trained at the inlet). Otherwise, the results can deteriorate, even for small distances between training set and measurement (this situation is given for case 2 with model variant lh, trained just ca. $2 \, \unit{cm}$ apart), most likely since the spatial variation is missing in the NN. Conversely, using multiple training positions in addition to the exact measurement location does not bring marked benefits, suggesting marginal transfer learning from the curves at different positions.

As found before at the phase-shifted measurements, the NN's precision in time is more critical for calibrating $C$, due to its time-dependent effect on the pressure. This carries over to coarsening the time grid, where lowered resolution affects $C$ more than $R$. Moreover, also the general trend observed above, that $R$ typically is recovered in a better way than $C$, is confirmed in these tests. Lastly, Case $1$, where the true parameters are mostly included in the training set, consistently performs better than Case $2$, where the true parameters are not contained in the training set. This indicates a solid interpolative capacity of the approach.

To sum up, the results reveal that the calibration remains reliable even when the training data are substantially reduced, where up to 4-5 times coarser meshes in each variable are fully admissible. This enables remarkable computational savings for generating new reference data.

\section{Conclusions and outlook}
\label{sec:conclusions}

We demonstrated that our neural network based calibration scheme provides a fast and accurate calibration of the global Windkessel parameters, even under moderate uncertainties in the reference measurement. Once learned the blood pressure at the left brachial artery based on simulations using a coupled 1D-0D model, the NN finds the global resistance and compliance parameters which best match a given measurement pulse wave. A key advantage of our approach is its automatic identification of the measurement location and the phase synchronization. The determined global WK parameters fully parameterize the 1D-0D model of the studied cardiovascular network, under our assumptions on the typical geometrical and mechanical vessel properties of the vessel, which allows for patient-specific simulation of blood flow, for example in the liver.

The first part of our research concerned the construction of a data-driven fully-connected neural network to replicate the map from WK parameters, measurement position and time to brachial blood pressure on the level of a 1D-0D simulation. We found that an architecture with two ReLU layers and a $\tanh$ layer, each with 32 neurons, could approximate the simulation data with an mean absolute error of at most $0.02 \, \unit{mmHg}$ per pulse wave. This architecture in particular captured the kinks in the pressure profile. We also confirmed that the NN's predictions are stable even with markedly coarser resolved training data, providing insight about the extent of the validity of a purely data-driven approach. Lastly, we described how to extend the neural network by dummy neurons, representing the calibration variables. This novel approach allows the calibration to be performed entirely within the NN framework by retraining the pretrained NN solely on these dummy neurons -- a process mathematically equivalent to least-squares optimization.

The second part analyzed the actual NN-based calibration to a reference measurement for various test cases. It was demonstrated, that the NN can correctly identify the WK parameters associated with one of its inputs curves, accurately correct phase mismatches and reliably determine the position of the measurement if unknown -- for moderately imprecise position in the NN, the calibration results kept stable as well. Moreover, our tests showed that the calibration tolerates significantly coarser resolution in the training data and still can yield accurate results with slightly distorted measurements.

The three main limitations of our approach concern the focus on global instead of individual WK parameters, its transferability and its high demands on measurement precision to identify the compliance parameters. Regarding the restriction to global WK parameters, it is clear that this necessarily misses to account for differences among the individual parts of the 1D-0D model, but results in an overall average cardiovascular model, based on typical vessel properties. While it would be desirable to calibrate the individual parameters instead, this would massively increase the dimensions of the parameter space to be covered by representative training data. Since this would also markedly scale up the training procedure, this seems infeasible within our frame-work.

Regarding the transferability, we note that the surrogate NN is very specific to the considered arterial system, i.e. the selection of 1D modeled vessels, the applied constants for vessel properties and the studied range of the WK parameters. We do not expect any variant of the NN trained on this system would provide reliable calibration results if the system assumptions are invalid. While we estimated the required resolution in new reference data, this data in fact must be generated first, if working on different systems. Moreover, while our idealizing numerical tests with synthetic data provide a solid proof of principle, it is unclear how this transfers to more realistic scenarios.

Regarding the recovery of the compliance, we observed that it requires for very precise pressure data in the time-resolved measurement. While the resistance parameter proved to be very robust to symmetric noise, the compliance parameter is affected by significant unsystematic error even at small noise strength. However, this seems to follow mainly from the specific nature of compliance on the pulse wave which is very subtle and time-depending, and even more, to some extend similar to the effect of the position. Even if our NN approach principally proved to relate these details in to the compliance, it might be very difficult to actually measure these effects. Thus, if only a few data are available, especially without time-resolution, it might be sufficient to calibrate only the resistance by the newly proposed method and to determine the compliance by means of the method described in Subsection \ref{sec:omitted}. Alternatively, it should be figured out if a different measurement setup, i.e., recording transient effect after clamping, is more informative for compliant effects.

Our future work will be concerned with enhancing the presented blood flow model by a FSI model considering viscoelasticity \cite{wang2015verification}. If there are data on the location of the blood vessels in space, gravitational effects can be included into the model \cite{shramko2023gravity}. Important mechanisms having a strong influence on the Windkessel parameters are auto-regulation mechanisms, which adjust the resistances of the arterioles such that there is a sufficient supply of the organs with blood and a sufficient removal of carbon dioxide from the interstitial space \cite[Chapter 2.4.2]{koppl2015multi}\cite{alastruey2008reduced}. The neural network that has been constructed to predict the blood pressure in the left brachial artery will be extended such that we can predict the blood pressure and flow rate in a complete network. For this purpose, it is required to estimate the parameters of a mathematical model emulating the heart beats. To estimate the duration of the heart beats, we plan to integrate data measured by an in-ear sensor \cite{zellhuber2024transforming}. However, it remains to estimate further parameter such as the maximal flow rate in \eqref{eq:Qheart} or the stroke volume of the heart. Regarding the outlets of the network, we intend to calibrate not only the Windkessel parameters of the whole arterial vessel system, but also Windkessel parameters of individual outlets. Apart from WK parameters, also E-moduli and wall thicknesses are crucial for the shape of pressure curves. These parameters depend significantly on the age of a human and several other factors. As a consequence, a patient specific calibration of these parameters is crucial. In addition to that, it would be interesting to integrate Doppler-based measurements to calibrate the model parameters. Finally, stochastic calibration methods could be considered to determine the model parameters \cite{richter2025bayesian}.

\subsection*{Supplementary Notebook}
The main features of the neural network can be retraced interactively on a Jupyter notebook uploaded to \href{https://github.com/bhoock/WKcalNN.git}{github.com/bhoock/WKcalNN.git}.

\subsection*{AI policy}
The authors used AI to improve the grammar, structure and readability during manuscript preparation. The final content was reviewed and approved by the authors which are exclusively responsible for the complete text.

\subsection*{Acknowledgements}
The authors gratefully acknowledge Luigi Iapichino for providing access to the computational and data resources of the Leibniz Supercomputing Centre (\href{www.lrz.de}{www.lrz.de}).
\section{Appendix: Choice of the model parameters}
\label{sec:model_parameters}
In this section, we list the model parameters for the blood flow model described in Section \ref{sec:bloodflow}. Table \ref{tab:simulation_parameters} contains the parameters for the boundary conditions as well as some material parameters. The other table (Table \ref{tab:DataNetwork}) is a summary of the network data. 
\begin{table}[h!]
	\begin{tabular}{|c|c|c|c|}
		\hline 
		Parameters & Value & Unit & Source \\
		\hline 
		blood density $\rho$ & $1.050 \cdot 10^{-3}$ & $\unitfrac{kg}{cm^3}$ & \cite{alastruey2007modelling}\\
		\hline
		venous pressure $p_v$ & $5.0$ & \unit{mmHg} & \cite{alastruey2007modelling} \\
		\hline 
		Poisson ratio $\nu$ & $0.5$ & \unit{---} & \unit{---}  \\
		\hline
		Duration of a heart beat $T$ & $60/70$ & \unit{s} & \cite{viitasalo1992qt}\\
		\hline
		Maximal flow rate $Q_{max}$ & $539.0$ &\unitfrac{$cm^3$}{s} & \unit{---} \\
		\hline
	\end{tabular}
	\caption{Parameters used in the simulation.}
	\label{tab:simulation_parameters}
\end{table}
For some parameters in Table \ref{tab:simulation_parameters} no source or reference is provided. This is the case, if these parameters are not directly taken from some reference or if some additional explanations are required. Due to the fact that biological tissue is practically incompressible, the Poisson ratio $\nu$ of the vessel walls is given by $0.5$. However, one has to be aware that this is in general a simplifying assumption, since biological tissue exhibits a complex structure. A precise and tissue specific measurement of $\nu$ requires high effort \cite{pajic2025compressibility}. In order to compute the maximal flow rate $Q_{max}$ that is used in \eqref{eq:Qheart}, we assume that there are $70$ heart beats per minutes such that $T$ is given by $60/70\;\unit{s}$. Moreover, a relatively high stroke volume $V_{st}$ of $100.0\;\unit{cm^3}$ is considered \cite{sidebotham2007physiology}. By this $Q_{max}$ can be determined using the following equation:
$$
V_{st} = \int_{0}^{T}\;Q_1\left(0,t\right) \;dt = Q_{max} \int_{0}^{\frac{T}{3}}\;\sin\left(\frac{\pi \cdot 3}{T}t\right)\;dt = \frac{2\cdot Q_{max} \cdot T}{\pi \cdot 3} \Leftrightarrow Q_{max} = \frac{3\pi \cdot V_{st}}{2 \cdot T}.
$$
\begin{table}[h!]
	\begin{center}
		\caption{\label{tab:DataNetwork} Data for the network in Figure \ref{fig:Network} (partially taken from \cite{stergiopulos1992computer,wang2004wave}).}		
		\begin{tabular}{|c|c|c|c|c|}
			\hline
			Vessel & Length    & Radius                 & Wall thickness          & E-modulus  \\
			$i$ & $l_i\;\left[\unit{cm}\right]$ & $R_{0,i}\;\left[\unit{cm}\right]$ & $h_{0,i}\;\left[\unit{cm}\right]$ & $E_i\;\left[10^6 \cdot \unit{Pa}\right]$ \\
			\hline
			$1$  & $4.0$ & $1.470$ & $0.163$ & $0.4$  \\
			\hline
			$2$  & $2.0$ & $1.120$ & $0.130$ & $0.4$  \\
			\hline
			$3$  & $3.4$ & $0.620$ & $0.080$ & $0.4$  \\
			\hline
			$4$  & $3.4$ & $0.423$ & $0.063$ & $0.4$ \\
			\hline
			$5$  & $17.7$ & $0.370$ & $0.045$ & $0.4$\\
			\hline
			$6$  & $14.8$ & $0.185$ & $0.045$ & $0.8$ \\
			\hline
			$7$  & $42.2$ & $0.310$ & $0.067$ & $0.4$ \\
			\hline
			$8$  & $17.6$ & $0.382$ & $0.045$ & $0.8$ \\
			\hline
			$9$  & $17.6$ & $0.382$ & $0.042$ & $0.8$ \\
			\hline
			$10$ & $3.9$  & $1.120$ & $0.115$ & $0.4$ \\
			\hline
			$11$ & $20.8$ & $0.370$ & $0.063$ & $0.4$ \\
			\hline
			$12$ & $17.6$ & $0.334$ & $0.045$ & $0.8$ \\
			\hline
			$13$ & $17.6$ & $0.334$ & $0.042$ & $0.8$ \\
			\hline
			$14$ & $5.2$ & $1.120$ & $0.110$ & $0.4$ \\
			\hline
			$15$ & $3.4$ & $0.474$ & $0.066$ & $0.4$ \\
			\hline
			$16$ & $14.8$ & $0.203$ & $0.045$ & $0.8$ \\
			\hline
			$17$ & $42.2$ & $0.310$ & $0.067$ & $0.4$ \\
			\hline
			$18$ & $8.0$ & $0.317$ & $0.049$ & $0.4$ \\
			\hline
			$19$ & $10.4$ & $1.017$ & $0.100$ & $0.4$ \\
			\hline
			$20$ & $5.3$ & $0.920$ & $0.090$ & $0.4$ \\
			\hline
			$21$ & $2.2$ & $0.350$ & $0.064$ & $0.4$ \\
			\hline
			$22$ & $2.3$ & $0.350$ & $0.064$ & $0.4$ \\
			\hline
			$23$ & $6.6$ & $0.350$ & $0.049$ & $0.4$ \\
			\hline
			$24$ & $5.3$ & $0.150$ & $0.045$ & $0.4$ \\
			\hline
			$25$ & $2.3$ & $0.100$ & $0.054$ & $0.4$ \\
            \hline			
		\end{tabular}
	\end{center}
\end{table}

\bibliography{literature_calibration_final}
\bibliographystyle{siam}

\end{document}